\documentclass[sigconf]{acmart}

\acmConference[ICSE 2022]{The 44th International Conference on Software Engineering}{May 21–29, 2022}{Pittsburgh, PA, USA}

\AtBeginDocument{%
  \providecommand\BibTeX{{%
    \normalfont B\kern-0.5em{\scshape i\kern-0.25em b}\kern-0.8em\TeX}}}

\copyrightyear{2022}
\acmYear{2022}
\setcopyright{acmcopyright}\acmConference[ICSE '22]{44th International Conference on Software Engineering}{May 21--29, 2022}{Pittsburgh, PA, USA}
\acmBooktitle{44th International Conference on Software Engineering (ICSE '22), May 21--29, 2022, Pittsburgh, PA, USA}
\acmPrice{15.00}
\acmDOI{10.1145/3510003.3510154}
\acmISBN{978-1-4503-9221-1/22/05}

\usepackage{booktabs}
\usepackage{multirow}
\usepackage{listings}
\usepackage{xcolor}
\usepackage{bm}

\definecolor{dkgreen}{rgb}{0,0.6,0}
\definecolor{gray}{rgb}{0.5,0.5,0.5}
\definecolor{mauve}{rgb}{0.58,0,0.82}

\usepackage{algorithmic}
\usepackage{graphicx}
\usepackage{textcomp}
\usepackage{xcolor}

\begin{document}

\title{Learning to Recommend Method Names with Global Context}


\author{Fang Liu}
\affiliation{%
  \institution{Key Lab of High Confidence Software \\ Technology, MoE (Peking University)}
  \city{Beijing}
  \country{China}
}
\email{liufang816@pku.edu.cn}

\author{Ge Li}
\authornote{Corresponding authors.}
\affiliation{%
  \institution{Key Lab of High Confidence Software \\ Technology, MoE (Peking University)}
  \city{Beijing}
  \country{China}
}
\email{lige@pku.edu.cn}

\author{Zhiyi Fu}
\affiliation{
  \institution{Key Lab of High Confidence Software \\ Technology, MoE (Peking University)}
  \city{Beijing}
  \country{China}
}
\email{ypfzy@pku.edu.cn}

\author{Shuai Lu}
\affiliation{
  \institution{Key Lab of High Confidence Software \\ Technology, MoE (Peking University)}
  \city{Beijing}
  \country{China}
}
\email{lushuai96@pku.edu.cn}

\author{Yiyang Hao}
\affiliation{
  \institution{Silicon Heart Tech Co., Ltd}
  \city{Beijing}
  \country{China}
}
\email{haoyiyang@nnthink.com}

\author{Zhi Jin}
\authornotemark[1]
\affiliation{%
  \institution{Key Lab of High Confidence Software \\ Technology, MoE (Peking University)}
  \city{Beijing}
  \country{China}
}
\email{zhijin@pku.edu.cn}

\renewcommand{\shortauthors}{Liu, et al.}


\begin{abstract}
In programming, the names for the program entities, especially for the methods, are the intuitive characteristic for understanding the functionality of the code. To ensure the readability and maintainability of the programs, method names should be named properly. Specifically, the names should be meaningful and consistent with other names used in related contexts in their codebase. In recent years, many automated approaches are proposed to suggest consistent names for methods, among which neural machine translation (NMT) based models are widely used and have achieved state-of-the-art results. However, these NMT-based models mainly focus on extracting the code-specific features from the method body or the surrounding methods, the project-specific context and documentation of the target method are ignored. We conduct a statistical analysis to explore the relationship between the method names and their contexts. Based on the statistical results, we propose GTNM, a Global Transformer-based Neural Model for method name suggestion, which considers the local context, the project-specific context, and the documentation of the method simultaneously. Experimental results on java methods show that our model can outperform the state-of-the-art results by a large margin on method name suggestion, demonstrating the effectiveness of our proposed model.
\end{abstract}






\maketitle

\section{Introduction}
During programming, developers must name variables, functions, parameters, \textit{etc}. The appropriateness of a name changes over time during the software evolution. For example, a good function name can degrade into a poor one when the semantics of the function change or the function is used in a new context. Poor names make programs harder to understand and maintain \citep{lawrie2006s,liblit2006cognitive,takang1996effects,arnaoudova2014repent,arnaoudova2016linguistic,hofmeister2017shorter}, leading to misuses and defects \citep{butler2009relating,abebe2011effect,abebe2012can,amann2018systematic}. Finding consistent names for program constructs has always been a cynosure in the software industry. 

Methods are the most minor named units for indicating the program behavior in most programming languages \citep{host2009debugging}, thus they are particularly important \citep{beck2007implementation,mcconnell2004code,martin2009clean}. Meaningful and conventional method names are vital for developers to understand the behavior of programs or APIs. Once the name of a method is decided, it is laborious to change, especially when used for an API \citep{allamanis2015suggesting}. The results from an investigation in \citet{liu2019learning} indicate that among the change history in projects, developers usually change the method names without any change to the corresponding body code in many cases, which suggests that programmers strive to choose meaningful and appropriate method names, i.e., more consistent with other names in the same project or the codebase. Especially when collaborating, they need to obey a project’s coding conventions. 

In recent years, researchers have proposed automated approaches for suggesting consistent names for those methods. Based on the intuition that two methods implemented with similar code in their body code are likely to be named similarly, \citet{liu2019learning} proposed an IR-based approach to detect and rename inconsistent method names. They identify the inconsistent method names by comparing the names retrieved from the method body vector space with those retrieved from the method name vector space. For the inconsistent names, their model recommends the potentially consistent names by referring to the names of similarly implemented methods. However, in many cases, even the methods with similar body code can be named differently because they might belong to different projects and have different semantics. Besides, by retrieving names from similar methods, the model cannot suggest neologisms.
\citet{allamanis2016convolutional} proposed a convolutional attentional network to extract local time-invariant and long-range topical attention features in the method body to suggest names for methods. To leverage the syntactic structure of programming languages, Code2vec \citep{alon2019code2vec} and Code2seq \citep{alon2019code2seq} represent the method body as a set of compositional abstract syntax tree (AST) paths and use the path representation to predict the method’s name. \citet{nguyen2020suggesting} proposed MNire, a simple but effective approach to recommend a method name and detect method name inconsistencies. They treated the method name generation task as an abstractive summarization of the tokens of the program entities' names in the method body and the enclosing class name. \citet{li2021context} developed DeepName, a context-based approach for method name consistency checking and suggestion. They extract the features from four contexts: the internal context, the caller and callee contexts, sibling context, and enclosing context.

The above state-of-the-art models mainly focus on exploiting code-specific features from the method body or the surrounding methods in the same program file, which can be considered as local contexts of a method. However, the information of the whole project (global context) is ignored in these models. For example, the documentation of the method can describe the method's functionality and the role it plays in the project. Besides, there also exist nested scopes for project, where 
a source code file can have references to other files of the same projects. Thus, the contexts from other program files which are imported by the file where the target method in are also helpful in understanding the methods. Intuitively, these contexts are of great importance for method name recommendation, especially for the methods which have little content in the body, but with sufficient global contexts. A method does not exist in isolation, a large number of associations can be found among the project-specific contexts and the documentation: (1) The functionality and naming convention of a method can be better understood when more contextual features are provided. (2) There might be many possible names that can match the semantic of the method. By referring to the global contextual information, the solution space of the method names can be narrowed. Thus, when recommending a method name, it is necessary to refer to the global contexts. It can help in following situations: when the method is first created, existing global context can be accessed for suggesting a proper name for it; during the code refinement, the global context can be used to suggest an alternative name if the current name is inconsistent.

To verify our intuition, we first conducted a statistical analysis to learn the relation between the method names and their contexts of different levels. Based on the statistical analysis results, we propose \textbf{GTNM}, a novel \textbf{G}lobal \textbf{T}ransformer-based \textbf{N}eural \textbf{M}odel for method name suggestion, aiming at generating meaningful and consistent names for methods. We treat the method name suggesting task as the abstractive text summarization, where \textit{the tokens from the contexts of different levels} are considered as input, and \textit{the sub-tokens in the method's name} is considered as the target summary of input sequences. We use the attention mechanism to allow the model to attend to different contexts during the decoding process. 

The main contribution of our model can be summarized as follows:
\begin{itemize}
    \item We conduct a statistical analysis to explore the relationship between the method names and their contexts of different levels.
    \item We propose a novel global approach for method name suggestion, which considers the local context, the project-level context, and the documentation of the method simultaneously.
    \item We conduct extensive experiments to evaluate our approach on the large-scale datasets of Java methods. The experimental results show that our model substantially improves the performance of the previous approaches on suggesting method names.
\end{itemize}

\section{Motivating Example and Statistical Analysis}
According to \citet{nguyen2020suggesting}, the principle of naturalness of software \citep{hindle2016naturalness} also holds for the tokens composing the names of program entities. Specifically, tokens are repetitive and occur in regularity, where the repetitiveness can be captured by statistical models trained on a large code corpus. Therefore, the tokens composing the names of program entities can reflect the semantic and functionality of the code snippets. Based on this evidence, most previous work mainly considers the associations among the tokens of the method names and the tokens in the method body (local context). However, only considering the local context is not sufficient. We assume that the project-specific context can better reflect the role that the target method plays in the whole project. For example, the methods in the same file with the target method (we call them in-file contextual methods) and the methods in other program files of the same project that are imported by the file where the target method locates (we call them cross-file contextual methods). Besides, the documentation of the method also plays an important role in recommending the method names. We present several java method examples to illustrate the associations among method names and the project-specific and documentation contexts in Section \ref{motivating example}, appearing as the token overlapping. Based on those observations, we conduct a statistical analysis to explore the relationship between the method names and the contexts of different levels in Section \ref{statistical analysis}, i.e., local context, project-specific context, and documentation context. 

\subsection{Definitions}

Firstly, we give a brief definition of tokens, local context, project-specific context, and documentation context.

\noindent\textbf{Definition of Tokens.} ~ For programs, we parse the program to AST and extract entities (method names, identifiers, parameters, return-types) from AST. Then we split the entities following camelcase and underscore naming conventions, and lowercase the entities to get tokens. For documentation, we extract the first sentence in Javadoc by deleting the punctuations. Then we split the sentence with space to get words and lowercase the words to get tokens.

\noindent\textbf{Definition of Local Context.} ~ Local-context contains the program entities in the method signature and body, including parameters, return type, and identifiers. 

\noindent\textbf{Definition of Project-specific Context.} ~
Project-specific context is supposed to reflect the target method's role in the whole project and the naming styles. We argue that the methods in the same file with the target method (we call them in-file contextual methods) and the methods in other program files of the same project that are imported by the file where the target method (we call them cross-file contextual methods) in can provide the above information. We consider the name of the contextual methods as the project-specific context. 

\noindent\textbf{Definition of Documentation Context.} ~
The first sentence of the code documentation is informative, and many code summarization approaches use it as a code summary \citep{hu2018deep,leclair2019neural,wei2019code}. Following them, we use the tokens in the first sentence as the documentation context. 

\subsection{Motivating Example}\label{motivating example}

1. The project-specific context might contain the entities that can provide semantic information for the target method name recommendation. In Code 1, the names of the third method (getMaxValue) do not describe the functionality of the methods well. When changing it into a more precise name that contains the project-related entity names (getMaximumResourceCapability), only referring to the method body is not enough. If the (in-file) project-level contextual information, i.e., other methods in the same file, can be accessed, we can easily realize that the method is related to the \textit{resource capability} and make correct revisions.

\begin{lstlisting}[caption={Project-specific context contains the entities that can provide semantic information}]
public Resource getClusterResource() {
    return clusterResource;
}
public Resource getMinimumResourceCapability() {
    return minimumAllocation;
}
// consistent name: getMaximumResourceCapability
public Resource getMaxValue() {
    return maximumAllocation;
}
\end{lstlisting}\label{code1}
\vspace{-0.1cm}

2. The project-specific contextual information can imply the logic and the functionality of the project, which will reflect the role the target method plays in the project. In Code2, these methods are related to the window events, including the keypress events or trackpad touch events. By accessing the (in-file) project-level context, the functionality of the whole project and the role of the target method can be better understood, thus offering more knowledge for recommending meaningful method name.

\begin{lstlisting}[caption={project-specific contexts imply the logic and the functionality of the project.}]
public boolean touchDown (InputEvent event, float x, float y, int pointer, int button) {
	...
}
public void touchUp (InputEvent event, float x, float y, int pointer, int button) {
	...
}
public boolean keyDown (InputEvent event, int keycode) {
	return isModal;
}
public boolean keyUp (InputEvent event, int keycode) {
	return isModal;
}
\end{lstlisting}
\vspace{-0.1cm}

3. There might be many semantically consistent names that can reflect the function of a specific method. We can narrow the solution space and suggest a consistent and conventional method name by referring to the project-specific contextual information. Both of the two methods in Code3 indicate that some errors are encountered. However, different verbs are used in the names (``Encountered'' and ``Occured''), and they are synonyms. Although these two names are both semantically correct, they are not consistent. When refactoring the second method name ``serverErrorOccured'' into a name that is consistent with the contextual methods, we can use the verb ``Encountered'' to replace ``Occured'' by referring to the previous method name ``clientErrorEncountered''. This suggests that with the help of the project-level context, we can choose the candidate names from a smaller and specific solution space.

\begin{lstlisting}[caption={Semantically consistent names}]
  public void clientErrorEncountered() {
    clientErrors.incr();
  }
  // consistent name: serverErrorEncountered
  public void serverErrorOccured() {
    serverErrors.incr();
  }
\end{lstlisting}  
\vspace{-0.1cm}

4. Cross-file project-specific context can provide extra information when the in-file context is less informative. In code4, the AccountActivity class inherits from BaseActivity class, thus the methods of the parent class BaseActivity might be overridden in AccountActivity class, for example, getLayoutRes(), onCreateActivity(), etc. The program file where the BaseActivity class defined is imported at the beginning of the file. Thus, we can extract the methods defined in the BaseActivity class by considering the cross-file project-specific contexts. Thus, when predicting the method name for the methods in AccountActivity class, the methods defined in its parent class can be accessed, which are helpful for the cases where the in-file context is less informative for inferring the method name. 

\begin{figure*}[] 
\setlength{\abovecaptionskip}{-0.2cm} 
\centering\includegraphics[width=13cm]{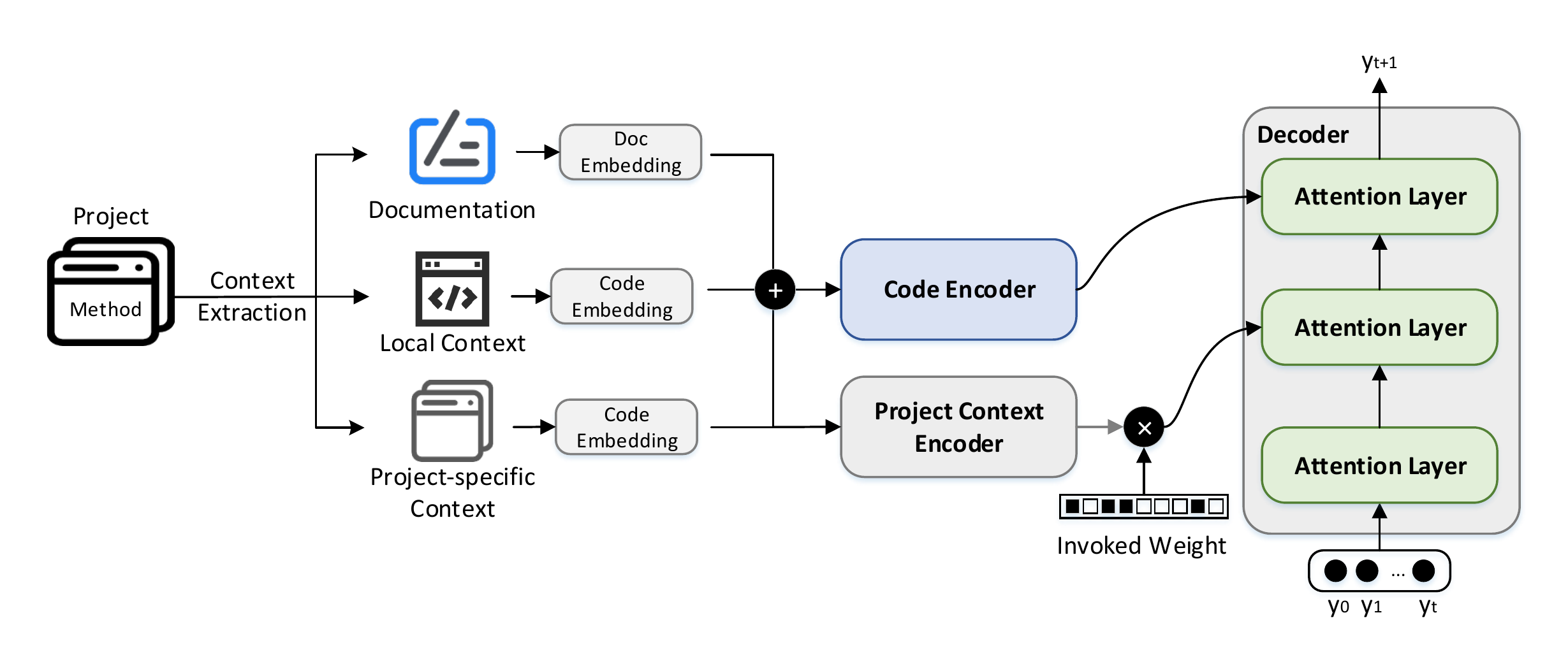} 
\caption{The overall framework of GTNM.}
\label{Fig:model_arch}
\vspace{-0.5cm}
\end{figure*}

\begin{lstlisting}[caption={Cross-file project-specific context can provide extra information when the in-file context is less informative}]
[AccountActivity.java]
...
import com.github.airsaid.accountbook.base.BaseActivity;
...
public class AccountActivity extends BaseActivity {

    @Override
    public int getLayoutRes() {
        return R.layout.activity_account;
    }
    @Override
    public void onCreateActivity(@Nullable Bundle savedInstanceState) {
        Account account = getIntent().getParcelableExtra(AppConstants.EXTRA_DATA);
        ...
    }
    ...
}
-----------------------------------------------------------------
[BaseActivity.java]
public abstract class BaseActivity extends SlideBackActivity {
    @Override
    protected void onCreate(@Nullable Bundle savedInstanceState) {
    ...
    }
    ...
    public abstract int getLayoutRes();
    public abstract void onCreateActivity(@Nullable Bundle savedInstanceState);
}
\end{lstlisting}\label{code5}
\vspace{-0.1cm}

5. The documentation can also provide rich information about the methods, which will help for suggesting method names. In Code5, the body code of these methods looks similar, and all of them cannot offer enough information for suggesting the method name. The documentation of the methods contains useful information that can reflect the functionality of the methods, thus being helpful for the method name recommendation. When predicting the name for the first method, the documentation can provide a useful indication. 

\begin{lstlisting}[caption={The documentation can provide rich information about the methods.}]
/**
 * Used to retrieve the plugin tool's descriptive name. */
// consistent name: getDescriptiveName
@Override
public String getName() {
    return "Remove Spurs (prunning)";
}
/**
 * Used to retrieve a short description of what the plugin tool does. */
@Override
public String getToolDescription() {
    return "Removes the spurs (prunning operation) from a Boolean image.";
}
\end{lstlisting} 
\vspace{-0.3cm}

\subsection{Statistical Analysis}\label{statistical analysis}
Based on the above observations, we conduct a statistical analysis to explore the relationships between the method names and their contexts by computing the percentage of their token sharing. For the analysis, we used the java programs in the Java-small dataset used in \citet{alon2019code2seq}. The dataset contains 11 high quality open-source java projects, which is a benchmark dataset for method name suggetstion task. It contains about 700K Java method examples. Thus, we use this dataset to conduct the statistical analysis to explore the relationships between the method names and their contexts. The statistical results in this analysis can be expected in a good project where most of the names are consistent.

For local context, we found that the tokens of 67.47\% of the method names can be found in the identifiers, and 35.64\% can be found in the return type and parameters. For Project-specific context, we found that the tokens of 85.98\% of the method names can be found in the names of its in-file contextual methods, and the tokens of 53.83\% of the method names can be found in the names of its cross-file contextual methods. For the documentation context, we found that the tokens of 55.98\% of the method names can be found in its documentation. There exists overlapping among different contexts, for example, the subtokens of the method name can appear in both local and documentation contexts. Thus, the sum of these numbers is not 100\%. Besides, 10.87\% of the method names cannot found in the body, but occur in the names of its project-specific context (in- and cross-file contextual methods). These results demonstrate that developers always refer to the project-specific context when naming the methods. Thus, project-specific context also contains essential information for method name recommendation, which should be carefully considered.

\begin{figure*}[t] 
\setlength{\abovecaptionskip}{0cm} 
\centering\includegraphics[width=16cm]{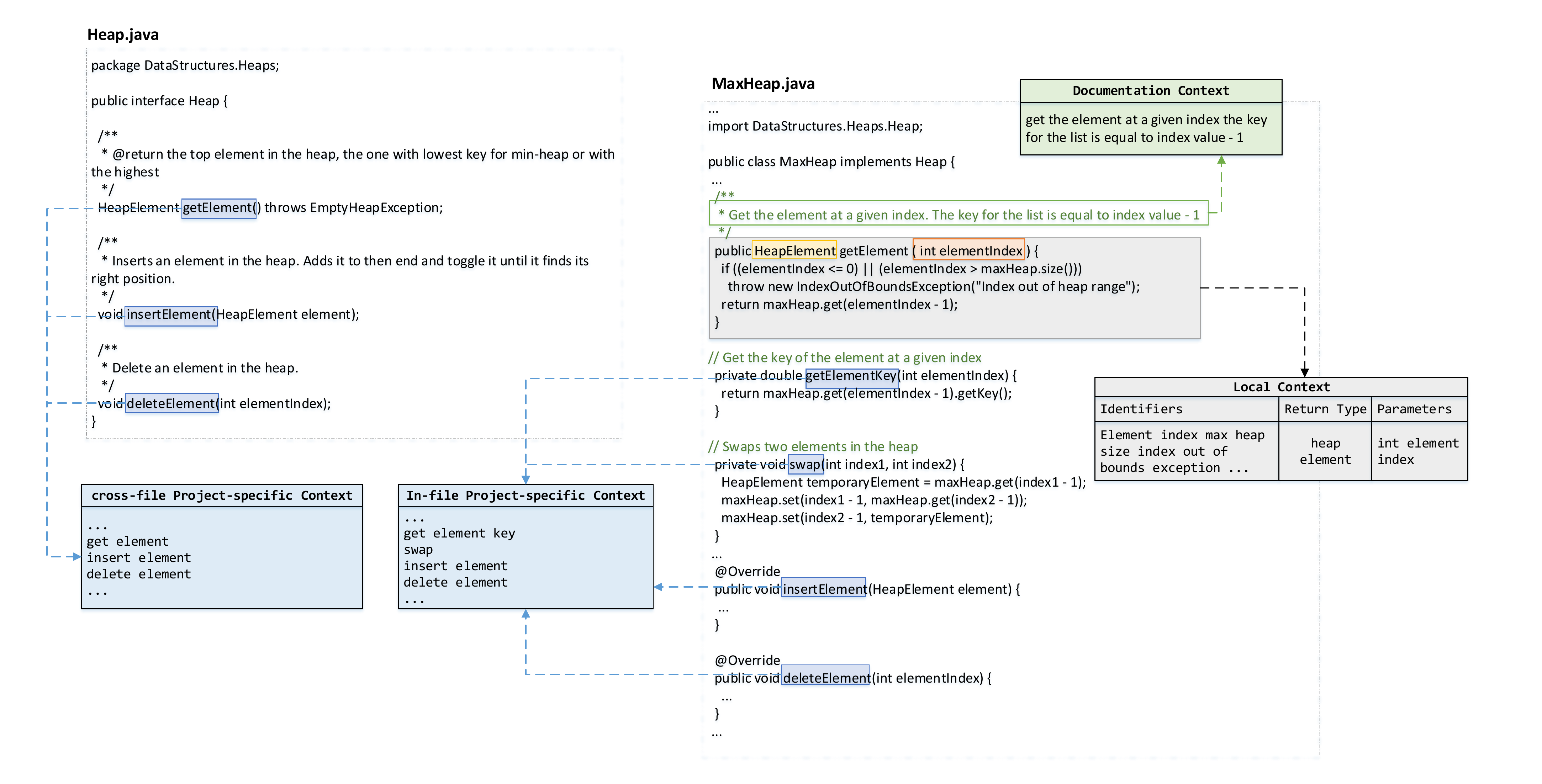} \caption{Different levels of contexts for method name suggestion.}
\label{Fig:cxt}
\vspace{-0.5cm}
\end{figure*}

\section{Proposed Model}

\subsection{Overview}
In this work, we propose GTNM, a global Transformer-based Neural Model for method name recommendation aiming at generating meaningful and consistent method names. The overall architecture of our approach is shown in Figure \ref{Fig:model_arch}. To fully utilize the contextual information of a method, we firstly extract context from three different levels given the target method and the project, including the local context, project-specific context, and documentation context. We employ a transformer-based seq2seq framework \citep{vaswani2017attention} to generate the method name. Specifically, we build corresponding encoders to encode the contexts into vector representations. The decoder generates the target method name by sequentially predicting the probability of the subtokens $y_{t+1}$ in the method name based on the contextual representations produced by the encoders, and the previous predicted subtokens $y1, y2, ..., y_t$. We use the attention mechanism to allow the model to attend to different contexts during the decoding process. 

\subsection{Context Extraction}
We extract the contexts of three different levels for generating meaningful and consistent names for the method, including local context, project-specific context, and documentation. Figure \ref{Fig:cxt} shows an example of the contexts for the Java method ``getElement''.

\noindent\textbf{Local Context Extraction}
According to the results of our statistical analysis and to represent the method body succinctly, we extract the following code entities as the local contexts for the method: (1) identifiers; (2) parameters; (3) return type. We tokenized each of the names from the local contexts following camelcase and underscore naming conventions, then normalized the tokens to lowercase. Finally, all the subtokens are concatenated in the order that they occurred in the source code to form the sequential representation of the local feature.

\noindent\textbf{Project-specific Context Extraction}
We define the project-specific context of one method as its in-file methods (other methods in the same file with the target method) and cross-file contextual methods (methods in the files imported by the file containing the target method). For simplicity and efficiency, we extract the name of the contextual methods as the project-specific context. Then we perform a similar process to these names as to local context. The concatenation of the lower-cased subtokens serves as the representation of the project-specific feature.

\noindent\textbf{Documentation Context Extraction}
For each method with a comment, to get its documentation context, we extract the first sentence that appeared in its Javadoc description since it typically describes the functionalities of the method\footnote{http://www.oracle.com/technetwork/articles/java/index-137868.html}. Then we delete the punctuations and split the sentence with space to get words and lowercase the words. All the words are concatenated to form the documentation context.

\subsection{Global Transformer-based Neural Model}
We use a transformer-based model to generate the method name, which leverages the self-attention mechanism and can capture rich semantic dependencies. The Transformer consists of stacked self-attention and point-wise, fully connected layers. The multi-head attention mechanism is performed in the self-attention layers. In each attention head, given the input vectors $\bm{x}=(\bm{x}_1,\bm{x}_2,...,\bm{x}_n)$, the output vectors $\bm{o}=(\bm{o}_1,\bm{o}_2,...,\bm{o}_n)$ is computed as:
\begin{equation}
    \begin{split}
        \bm{o}_i &= \sum_{j=1}^n \alpha_{ij} (\bm{x}_j\bm{W}^V)  \\
        \alpha_{ij} &= \frac{\text{exp}( e_{ij})}{\sum_{k=1}^n \text{exp}(e_{ik})} \\
        e_{ij} &= \frac{\bm{x}_i\bm{W}^Q(\bm{x}_j\bm{W}^K)^T}{\sqrt{d_k}}
    \end{split}
\end{equation}
where $\bm{W}^Q, \bm{W}^K \in \mathbb{R}^{d_{model} \times d_k}, \bm{W}^V \in \mathbb{R}^{d_{model} \times d_v}$ are the trainable parameters that are unique per layer and per attention head. Then the outputs of all the heads are concatenated to produce the final output of the self-attention layer.

After the attention layers of both encoder and decoder, a fully connected feed-forward network is employed:
\begin{equation}
    \begin{split}
        FFN(\bm{x}) = max(0, \bm{x}\bm{W}_1 + \bm{b}_1)\bm{W}_2 + \bm{b}_2
    \end{split}
\end{equation}
where $\bm{W}_1 \in \mathbb{R}^{d_{model} \times 4d_{model}}$, $\bm{W}_2 \in \mathbb{R}^{4d_{model} \times d_{model}}$, $\bm{b}_1 \in \mathbb{R}^{4d_{model}}$, $\bm{b}_2 \in \mathbb{R}^{d_{model}}$ are the trainable parameters.

\noindent\textbf{Encoders.}
We build a Code Encoder to encode the whole context $x$ including the local context, project-specific context, and documentation for the method name generation, and build an extra Project Context Encoder to encode the project context $x_{pro}$ for enhancing the attention to the project-specific context.

\noindent\textbf{i) Code Encoder.}
The local context, project-specific context and the documentation context are first embedded into vectors $\bm{x}_{loc}$, $\bm{x}_{pro}$, $\bm{x}_{doc}$, then these vectors are concatenated to form the representation of the whole contexts $\bm{x} = concat(\bm{x}_{loc}, \bm{x}_{pro}, \bm{x}_{doc})$, where $|\bm{x}| = |\bm{x}_{loc}|+|\bm{x}_{pro}|+|\bm{x}_{doc}|$. Then we employ transformer-based encoder to encode $\bm{x}$ into hidden representation $\bm{h}=(\bm{h}_1,\bm{h}_2,...,\bm{h}_{|\bm{x}|})$.

\noindent\textbf{ii) Project-specific Encoder.}
To increase the attention for the project-specific context, especially for the method names where the target method invoked, we build a Project-specific Encoder to encode the project-specific context $\bm{x}_{pro}$ into hidden representation $\bm{h}_{pro}=(\bm{h}_1^{pro},\bm{h}_2^{pro},...,\bm{h}_{|\bm{x}_{pro}|}^{pro})$. We use a mask vector $\bm{M} \in \mathbb{R}^{|\bm{x}_{pro}|}$ to record the methods that are invoked by the local context. $M_i$ is 1 if the $i$-th method in the project-specific context is invoked by the local context else is 0.

Intuitively, the methods in the project-specific context invoked by the local context are more important and relative to the target method. Thus we give these methods more attention by multiplying the invoked weight $\bm{w}$ on the project-specific hidden vector $\bm{h_{pro}}$ to produce the final project-specific hidden vector $\Tilde{\bm{h}}_{pro}$:
\begin{equation}\label{eq:weights}
\begin{split}
    &\bm{w} = softmax(1+\bm{M}) \\
    &\Tilde{\bm{h}}_{pro} = \bm{w} \otimes \bm{h}_{pro}
\end{split}
\end{equation}
where $\otimes$ is the element-wise production operation.

\noindent\textbf{Decoder.}
The decoder aims to generate the target method name by sequentially predicting the subtoken $y_{t+1}$ conditioned on the context vectors $\bm{h}$ and $\Tilde{\bm{h}}_{pro}$, and the previous generated subtokens $\bm{y}_{1:t}$:

\begin{equation}
\begin{split}
    &p(y_{t+1}) = \text{softmax}(FFN(\bm{dec}_2)) \\
    &\bm{dec}_2 = \text{Attention-Layer3}(\bm{h}, \bm{dec}_1) \\
    &\bm{dec}_1 = \text{Attention-Layer2}(\Tilde{\bm{h}}_{pro}, \bm{dec}) \\
    &\bm{dec} = \text{Attention-Layer1}(\bm{y}_{1:t})
\end{split}
\end{equation}
where the first attention layer performs multi-head attention over the decoder input $y_{1:t}$ to produce the hidden representation $\bm{dec}$. Then the second attention layer performs multi-head attention over the weighted project-specific hidden vector $\Tilde{\bm{h}}_{pro}$ to produce the hidden representation $\bm{dec}_1$, which models the dependency between the decoder input and the project-specific context. The last attention layer performs multi-head attention over the whole context hidden vector $\bm{h}$ to produce the final hidden representation $\bm{dec}_2$, which models the dependency between the decoder input, project-specific context, and the whole context. Then the final hidden representation is fed into a fully connected feed-forward network and softmax layer to produce the probability of the next subtoken $y_{t+1}$ for the target method name.

\noindent\textbf{Training.}
To train the network, we adopt cross-entropy loss between the predicted distribution $\bm{q}$ and the “true” distribution $\bm{p}$, which is computed as:
\begin{equation}
    H(\bm{p}||\bm{q}) = - \sum_{y\in Y} p(y)\log q(y) = -\log q(y_{true})
\end{equation}
where $y_{true}$ is the target name. Since p will assign value of 1 to the actual label in the training example and 0 otherwise, the cross-entropy loss for a example is equivalent to the negative log-likelihood of the true label. As $q(y_{true})$ tends to 1, the loss approaches zero. The smaller $q(y_{true})$ goes, the greater the loss becomes. Thus, minimizing this loss is equivalent to maximizing the log-likelihood that the model assigns to the true labels.

\begin{table}[t]
\centering
\setlength{\abovecaptionskip}{0.1cm} 
\caption{Statistics of the datasets.}
\begin{tabular}{lccc}  
\toprule
 ~ & Train & Validation & Test \\
\midrule
Files & 1,700,000 & 393,327 & 61,000 \\
Methods &  18,230,509 & 4,283,580 & 636,816\\
Methods with doc & 4,264,852 & 964,078 & 143,913 \\
\bottomrule
\end{tabular}
\vspace{-0.3cm}
\label{tab:datasets}
\end{table}

\section{Experimental Setup}
\subsection{Datasets}
We train and evaluate GTNM on Java programs following MNire \citep{nguyen2020suggesting} and Code2vec \citep{alon2019code2vec}. \citet{nguyen2020suggesting} provide the list of java repositories, which contains 10K top-ranked, public Java projects on GitHub. They used the same setting as in code2vec to shuffle files in all the projects and split them into 1.7M training and 61K testing files. Following their setting, we download the repositories they provide and follow the same way to build the dataset. After data processing, the detailed data information is shown in Table \ref{tab:datasets}.

\subsection{Metrics}
To evaluate the quality of the generated method name, we adopted the metrics used by previous works \citep{alon2019code2vec,alon2019code2seq,nguyen2020suggesting}, which measured \textit{Precision}, \textit{Recall}, and \textit{F-score} over sub-tokens. Specifically, for the pair of the target method name $t$ and the predicted name $p$, the $precision(t,p)$, $recall(t,p)$, and $F1(t,p)$ score are computed as:
\begin{equation}
    \begin{split}
        precision(t,p) &= \frac{|\rm{subtoken}(t)| \cap |\rm{subtoken}(p)|}{|\rm{subtoken}(p)|} \\
        recall(t,p) &= \frac{|\rm{subtoken}(t)| \cap |\rm{subtoken}(p)|}{|\rm{subtoken}(t)|} \\
        F1(t,p) &= \frac{2\times precision(t,p) \times recall(t,p)}{precision(t,p) + recall(t,p)} \\
    \end{split}
\end{equation}
where subtoken($n$) return the subtokens in the name $n$. Precision, Recall, and F-score of the set of the suggested names are defined as the average ones on all samples. Besides, we also measure the \textit{Exact Match Accuracy (EM Acc)}, in which the order of the subtokens are also taken into consideration.

\begin{table}[t]
\centering
\setlength{\abovecaptionskip}{0.1cm} 
\caption{Statistics of contexts and target name lengths.}
\begin{tabular}{lcc}  
\toprule
~ &  Avg & Med \\
\midrule
In-file Contextual Method  & 1399 & 68 \\
Cross-file Contextual Method & 197 & 80 \\
Variables & 23 & 7\\
Parameter and return type & 3 & 3 \\
Target Names & 3 & 2 \\
\bottomrule
\end{tabular}
\vspace{-0.3cm}
\label{tab:length}
\end{table}

\subsection{Implementation Details}
We use Transformer with 6 layers, hidden size 512, and 8 attention heads for both encoders and decoders. The inner hidden size of the feed-forward layer is 2048. We use javalang\footnote{https://github.com/c2nes/javalang} to parse the java code to extract the contexts. The details of different contexts and target names (subtoken) lengths are shown in Table \ref{tab:length}.  

In our experiments, we set the in-file project-specific context length to 30, the cross-file project-specific context length to 30, the local context length to 55 (variable length (50) + parameter and return type length (5)), the documentation context length to 10. And the maximum target name length is set to 5 \footnote{we examined model's performance with different context length settings, the setting that can achieve the best results were used for the final training}. We use the same vocabulary for the input source code and the target method name and build another vocabulary for the documentation context. The vocabulary size for the source code is set to 20,000, and the vocabulary size for documentation is set to 10,000. The out-of-vocabulary tokens are replaced by $\langle\texttt{UNK}\rangle$. To demonstrate the effectiveness of the cross-file project-specific context, we conduct experiments under the cross-project setting where the programs used in the training and test process are from different projects. Since more contexts can be accessed, we assume that we can use fewer programs to train the model. To verify the assumption, we train the model using a subset of the whole training dataset and compare it with the results without using the cross-file project-specific context. The detailed results are presented in \ref{RQ3}.

We use Adam with the learning rate of 3e-4, linear learning rate warmup schedule over the first 4,000 steps to train the model for 20 epochs. We use a dropout probability of 0.3 on all layers. Our model is trained on one Tesla V100 GPU with 16GB memory.

\section{Research Questions and Results}
To evaluate our proposed approach, in this section, we conduct experiments to investigate the following research questions:

\subsection{RQ1: Comparison against state-of-the-art models} 

We compare GTNM with the following state-of-the-art method name suggestion models:

\noindent 1) code2vec \citep{alon2019code2vec}: an attention-based neural model, which performs attention mechanism over AST paths and aggregates all of the path vector representations into a single vector. They considered the method name prediction as a classification problem and predicted a method’s name from the vector representation of its body.

\noindent 2) code2seq \citep{alon2019code2seq}: an extended approach of code2vec, which employs seq2seq framework to represent AST paths of the method body node-by-node using LSTMs and then attend to them while generating the target subtokens of the method name.

\noindent 3)  MNire \citep{nguyen2020suggesting}: an RNN-based seq2seq model approach to suggest a method name based on the program entities' names in the method body and the enclosing class name.

\noindent 4) DeepName \citep{li2021context}: an RNN-based approach for method name consistency checking and suggestion, using both internal and interaction contexts for method name consistency checking and suggestion, which achieves the state-of-the-art results on java method name suggestion task.

\begin{table}[t]
  \begin{center}
  \setlength{\abovecaptionskip}{0.1cm} 
  \caption{Method name recommendation comparison results.}
    \begin{tabular}{lcccc} 
     \toprule
      Model & Precision & Recall & F1 & EM Acc\\
      \midrule
      code2vec\citep{alon2019code2vec} &  51.93\% & 39.85\% & 45.10\% & 35.59\% \\
      code2seq\citep{alon2019code2seq} &  68.41\% & 60.75\% & 64.36\% & 41.50\% \\
      MNire\citep{nguyen2020suggesting} &  70.10\% & 64.30\% & 67.10\% & 43.10\% \\
      DeepName\citep{li2021context} & 73.60\% & 71.90\% & 72.70\% & 44.30\% \\
      \midrule
    %   GRU & 74.84\% & 56.21\% & 64.20\% & 35.52\% \\
      GTNM & \textbf{77.01\%} & \textbf{74.15\%} & \textbf{75.60\%} & \textbf{62.01\%} \\
      \bottomrule
    \end{tabular}
    \label{tab:gen_res}
    \vspace{-0.5cm}
  \end{center}
\end{table}

The first three baselines do not use the cross-file project-specific context for the method name suggestion. To make the comparison fair, we do not use the cross-file project context in this experiment. We use the same dataset as MNire and DeepName to train our model. For code2vec and code2seq, we download their publicly available source code and train their model on the same datasets. The results are shown in Table \ref{tab:gen_res}. Among these baselines, code2vec and code2seq only use the context in the method body to predict the method names. MNire utilizes the enclosing class (where the method is in) contexts, and DeepName further considers the interaction context and sibling context, which might appear in other program files. 

The results show that GTNM outperforms all the baseline models on all the metrics by a large margin, especially on the exact match accuracy. The higher exact match accuracy indicates the generated name is more close to the ground truth. Table \ref{tab:examples} shows two examples where the exact-match didn't occur but F1 was good. In the first case, the semantics of two names are reverse although they shared most of the sub-tokens with a high F1 score. Thus, exact match accuracy can evaluate the generated name more precisely, which plays a crucial role in method name suggestion. There are 32\% of the test methods where exact match is not satisfied but F1 $\geq$ 0.5. Among these cases, only 2.32\% of methods have the same subtoken set between generated names and target name.

\begin{table}[]
    \centering \small
    \setlength{\abovecaptionskip}{0.1cm} 
    \caption{Examples where the exact match did not occur but F1 was good.}
    \begin{tabular}{c|c}
    \toprule
     Prediction  &  Ground Truth\\
     \midrule
      `before', `attach', `primary', `storage'  & `before', `detach', `primary', `storage' \\
      `reset', `buffer' & `reset' \\
      \bottomrule
    \end{tabular}
    \vspace{-0.5cm}
    \label{tab:examples}
\end{table}

Although MNire and DeepName also consider the contexts beyond the method body, the contexts extracted by their approaches are different from ours. They only consider the contexts directly interacting with the target method, such as the sibling methods, callers methods, and callees methods. However, the methods which have no explicit interaction with the target methods can also provide essential information for understanding the functionality of the target method. For example, the methods appeared in the imported files, as shown in our previous motivation examples. Besides, MNire and DeepName use an RNN-based model to learn the relationship among the entities in the context. In our model, we extract contexts from a larger set of program entity candidates and employ a powerful backbone model to model the contexts, which is based on the self-attention mechanism. Besides, we also give the project-specific contexts more attention weights by applying invoked weight matrix. When generating the names of the target method, different decoder layers are utilized to focus on the contexts of different levels. Thus, our model can achieve better performance than baseline models.

Among these metrics, the exact match accuracy is much more strict than the other three metrics, which calculates the percentage of the predicted method names that are exactly the same as the ground truth. The other three metrics are based on the subtoken overlapping between the predicted names and the target names, where the order of the subtokens is ignored. 
The results show that our model obtains larger improvements on exact match accuracy and recall, which further demonstrates that the subtokens in the predicted names generated by our model can cover much more target subtokens than the other baselines. Therefore our model can fully and precisely describe the functionality of the method body.

\begin{table}[t]
  \begin{center}
  \setlength{\abovecaptionskip}{0.1cm} 
  \caption{Performance of using different contexts.}
    \begin{tabular}{lcccc} 
     \toprule
      Model & Precision & Recall & F1 & EM Acc \\
      \midrule
      Token seq & 70.25\% & 64.75\% & 67.39\% & 49.44\% \\
      Local cxt & 69.60\% & 64.38\% & 66.89\% & 50.95\% \\
      + In-file Project cxt & 75.16\% & 71.83\% & 73.46\% & 59.51\% \\
      + Documentation cxt & \textbf{77.01\%} & \textbf{74.15\%} & \textbf{75.60\%}& \textbf{62.01\%} \\
      \bottomrule
    \end{tabular}
    \label{tab:model_var}
    \vspace{-0.5cm}
  \end{center}
\end{table}

\begin{table}[]
  \begin{center}
  \setlength{\abovecaptionskip}{0.1cm} 
  \caption{The results on the extracted documented methods.}
    \begin{tabular}{lcccc} 
     \toprule
      Model & Precision & Recall & F1 & EM Acc \\
      \midrule
      GTNM & 85.36\% & 82.54\% & 83.93\% & 70.60\% \\
      - doc & 80.31\% & 76.65\% & 78.44\% & 64.14\% \\
      \bottomrule
    \end{tabular}
    \label{tab:doc_res}
    \vspace{-0.5cm}
  \end{center}
\end{table}

\subsection{RQ2: The contributions of contexts in the same file} 
In the previous experiment, we consider contexts of the same file (i.e., local context, in-file project-specific context, and documentation context) for generating the method name. To answer this research question, we conduct experiments using different context combinations. As shown in Table \ref{tab:model_var}, the first row shows the results of only taking the source code token sequence in the method body as input. The second row presents the result of using the local context (i.e., the entities' names of the method signature and variables) as input to suggest the method name. The third row shows the results of using both the in-file project-specific context and local context. The last row gives the results of using all three contexts: local, in-file project-specific, and documentation context. 

As seen from Table \ref{tab:model_var}, comparing the results of using local context (sequence length is 55) with the results of using source code token sequence (sequence length is 200), the performance is comparable, and using the local context can achieve higher exact match accuracy. However, the length of the local context is much shorter than the source code token sequence, which demonstrates that the local context extracted by our model contains enough information about the functionality of the method body, and the shorter context can improve the computational efficiency of the model. When we further incorporating the project-specific context information, the performance is improved by a large margin. Specifically, the F1 score and exact match accuracy significantly increase from 66.89\% and 50.95\% to 73.46\% and 59.51\%. The substantial improvement shows that the project-specific context, which can offer knowledge about the project information, is essential and efficient for improving the performance of method name recommendation. 

When the documentation information is added, the performance is further improved. However, in our whole dataset, only about 20\% of the methods have the document information. Thus, for most of the methods, the documentation context information is missing. To directly illustrate the contribution of the documentation context information, we extract those documented methods from the whole dataset and present the results on the extracted dataset. As shown in Table \ref{tab:doc_res}, the first row shows the results of our full model on the extracted dataset, and the second row shows the results of removing the documentation context from the input. When removing the documentation context, the performance is decreased by 5.1 in precision, 5.9 in recall, 5.5 in F1, and 6.5 in exact match accuracy, respectively. The results demonstrate that the documentation context can provide useful information for the method name suggestion.

\subsection{RQ3: The contribution of cross-file context}\label{RQ3}
When considering the cross-file project-specific context, we need to preserve the project structure of the programs in the dataset. Since more contextual information can be accessed, we assume that the model can be trained in a low-resource setting, that is, fewer programs are needed for training the model. Thus, we only use a subset of the whole training dataset in this experiment. Specifically, we sample 4000 projects from the big training set as a small training set and extract the cross-file project-specific context for the programs in the sampled projects. We compare with the results of our model setting without using project-specific context. To further demonstrate the effectiveness of the cross-file project-specific context, we conduct the experiment under the cross-project setting. That is, we split the corpus based on the projects instead of files or the methods. The cross-project setting is challenging and reflects better the real-world usage of the method name recommendation where the model is trained on the set of existing projects and used to check for a new project.

The results are shown in Table \ref{tab:pro_res}. As seen from the results, with the help of cross-file project-specific context, our model can achieve comparable results with the results of the previous model setting, where the training set is bigger and in-project split, only using less than 50\% of the whole training set and under the challenging cross-project experimental setting. When removing the cross-file project-specific context, the performance of the model drops a lot, which further demonstrates the importance of the cross-file project-specific context.

\begin{table}[t]
  \begin{center}
  \setlength{\abovecaptionskip}{0.1cm} 
  \caption{Performance of using cross-file project-specific context under cross-project and low-resource setting.}
    \begin{tabular}{lcccc} 
     \toprule
      Model & Precision & Recall & F1 & EM Acc \\
      \midrule
      w/o cross-file cxt &  67.25\% &  64.66\% &  65.93\% &  49.71\% \\
      w/ cross-file cxt &  73.52\% &  70.65\% &  72.06\% & 60.69\% \\
      \bottomrule
    \end{tabular}
    \label{tab:pro_res}
    \vspace{-0.5cm}
  \end{center}
\end{table}

\section{Discussion}

\lstset{frame=none}
 \begin{table*}
  \begin{center}
  \setlength{\abovecaptionskip}{0.1cm} 
    \caption{Examples of generated summaries given Java methods.}
    \setlength{\tabcolsep}{2mm}{
    \label{tab:case_study}
    \begin{tabular}{l|l} 
      \toprule
      \multicolumn{2}{c}{\bf{Examples}} \\
      \midrule
      Method 1  & 
      \begin{lstlisting}
/** 
 * Adds a path (but not the leaf folder) if it does not already exist. */ 
protected void ____ (List<String> path, int depth) 
{ 
    int parentSize = path.size() - 1; 
    String name = path.get(depth); 
    Folder child = getChild(name); 
    if (child == null) 
    { 
        child = new Folder(name); 
        ...
} 
      \end{lstlisting}
       
      \\
    \midrule
      Human-written & "add" \\
      GTNM & "add", "path", "if", "not", "exists" \\
    \bottomrule
    Method 2  & 
    \begin{lstlisting}
/**
 * Append the longs in the array to the selection, each separated by a comma */
private void ____ ( long[] objects ) {
    for ( int i = 0; i < objects.length; i++ ) {
        selection.append( objects[i] );
        if ( i != objects.length - 1 ) {
                selection.append( ',' );
        }
    }
}
    \end{lstlisting}
      \\
 \midrule
      Human-written  &  "join", "in", "selection" \\
      GTNM &  "append", "selection" \\
\bottomrule

    Method 3  & 
    \begin{lstlisting}
/**
 * Calculates the DefinitionUseCoverage fitness for the given DUPair on the 
 given ExecutionResult */
public double ____ () {
    if (isSpecialDefinition(goalDefinition))
        return calculateUseFitnessForCompleteTrace();
    double defFitness = calculateDefFitnessForCompleteTrace();
    if (defFitness != 0)
        return 1 + defFitness;

    return calculateFitnessForObjects();
}
    \end{lstlisting}
      \\
 \midrule
      Human-written  &  "calculate", "d", "u", "fitness" \\
      GTNM & "calculate", "fitness", "for" \\
\bottomrule

    Method 4  & 
    \begin{lstlisting}
/**
 * Validate removal of invalid entries. */
public void ____ () {
    RightThreadedBinaryTree<Integer> bt = 
    new RightThreadedBinaryTree<Integer>();
    assertFalse (bt.remove(99));
    bt = buildComplete(4);
    assertFalse (bt.remove(99));  
    assertFalse (bt.remove(-2));  
}
     \end{lstlisting}
     \\
    \midrule
      Human-written  &  "test", "invalid", "removals"  \\
      GTNM &  "test", "remove", "invalid" \\
\bottomrule

    \end{tabular}
    }
  \vspace{-0.3cm}
  \end{center}
\end{table*}

\subsection{Qualitative Analysis}\label{case}
We perform qualitative analysis on the human-written method names and method names which are automatically generated by GTNM. In most cases, the names generated by GTNM are exactly the same as the human-written names. To figure out in what cases our model generates different names with human, we randomly sample 200 cases where the names generated by our model are different from the ground truth from the test to analyze the results. 

Following \citet{mcburney2015automatic} and \citet{hu2020deep}, we performed qualitative analysis to obtain opinions from participants on the quality of the generated-name, aiming at getting the feedback on our approach and directions for future-work. We invited 8 volunteers with 3-5 years of Java development experience to evaluate the generated names of the sampled 200 cases in the form of a questionnaire. Each participant is asked to answer several questions, including whether the human-written-names or generated-names are good, what are the differences between two names, etc. According to the questionnaire results, we summarize top-4 representative situations (The proportion of each situation is 19.4\%/43.6\%/6.6\%/11.9\%) as shown in Table \ref{tab:case_study}.

\noindent\textbf{Contain More Detailed Information.} ~
As shown in method 1, human just names the method as ``add''. What and when to add is not given. The human-written method name is very short and cannot reflect the detailed role of the target method. In cases like this, GTNM tends to generate a longer name that contains more information about the method's functionality. In this example, GTNM suggests a more detailed name ``add path if not exists'', which indicates that the object and the usage scenario of the target method. Our model can learn this detailed information from the documentation, parameters, and the method body. In the whole test set, 25\% of the wrong cases belong to this situation.

\noindent\textbf{Synonyms.}  ~
As shown in method 2, the human-written name and the name generated by our model have the same meaning, and the verbs used in these two names are synonyms (``join in'' and ``append''). Since ``join in'' is not as often used as ``append'' in the method names, and the contexts (including the project-specific context, local context, and the documentation context) also do not offer the relevant information about it. Thus, GTNM cannot correctly suggest the subtokens ``join in''. However, the name generated by our model can also precisely describe the functionality of the target method, which is also semantic consistent and acceptable.

\noindent\textbf{Acronym.} ~
In method 3, the human-written name contains an acronym for the specific entities, i.e., ``du'' for ``definition use'', which our model cannot correctly infer. Based on the given contexts, GTNM suggests a name that has a similar style with the project-specific context, but fails to suggest the acronym for specific entity names. 

\noindent\textbf{Different Word Orders.}  ~
As shown in method 4, the subtokens in the human-written name and GTNM suggested name are almost the same (except for ``removals'' and ``remove''), but the subtoken orders are different. In this example, the different orders do not affect the semantic of the method name, and both of the two names express the same meaning. However, in other cases, the semantic of the names with different subtoken orders might be different. 0.7\% of the wrong cases belong to this situation.

\begin{figure}[t] 
\setlength{\abovecaptionskip}{0cm} 
\centering\includegraphics[width=8cm]{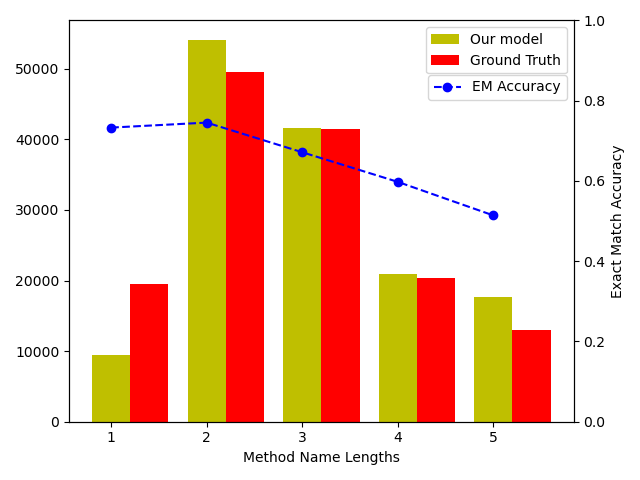}
\caption{The method name length distribution and the exact match accuracy of different name lengths}
\label{Fig:length_dis}
\vspace{-0.5cm}
\end{figure}

\subsubsection{Length analysis}
We further analyze the generated name length distribution and the performance of GTNM for different name lengths. As shown in Figure \ref{Fig:length_dis}, the lengths of the method names (the number of subtokens in the method name) mainly range from 2 to 3. Our model generates fewer names of length 1, and generated more names with lengths 4 and 5. Among all the methods, only 13.78\% of the names generated by our model are shorter than the ground truth. We apply the Wilcoxon Rank Sum Test (WRST) \citep{wilcoxon1992individual} to test whether the increase in the method name length is statistically significant, and all the p-values are less than 1e-5, which indicates a significant increase. We also use Cliff’s Delta \citep{macbeth2011cliff} to measure the effect size, and the values are non-negligible. Thus, our model tends to suggest more detailed names for the method. Besides, we also give the exact match accuracy of different lengths. As the length increase, the method naming task becomes harder. Even though our model can still achieve more than 50\% accuracy for the names of length 5.

\subsection{Explainability Analysis}

Lack of explainability is an important concern in many complex AI/ML models in SE \cite{pei2017deepxplore,tian2018deeptest}. It is crucial to ensure that the model is learned correctly and the logic behind the model is reasonable, which is also important for method name recommendation task. In this section, we analyze the explainability of GTNM. We employ model's confidence about its prediction to decide whether to accept the model's recommendation. Prediction Confidence Score (PCS) \cite{zhang2020towards} which depicts the probability difference between the two classes with the highest probabilities is a measure for evaluating model's confidence. In our model, the Pearson Correlation Score between PCS and F1-score of the generated names is 0.612 and p-value \textless 0.05, demonstrating that the correctness of the generated name is closely related to the model's confidence about its prediction. Thus, users can decide whether to accept the generated names depending on the case's error tolerance and the model's confidence.

\subsection{Threats to Validity}
\noindent\textbf{Threats to external validity} relate to the quality of the dataset we used and the generalizability of our results. We evaluate our approach on the Java dataset, which is a benchmark dataset for method name suggestion, and has been used in previous work \citep{alon2019code2vec,alon2019code2seq,nguyen2020suggesting}. All of the programs in the dataset are collected from top-ranked and popular GitHub repositories. Thus, most-of-the-names are expected consistent. However, there still exist a few cases that the name is inconsistent as shown in section \ref{case}. Besides, further studies are also needed to validate and generalize our findings to other programming languages. Furthermore, our case study is on a small scale. More user evaluation is needed to confirm and improve the usefulness of our model.

\noindent\textbf{Threats to internal validity} include the influence of the model architectural choices and the hyper-parameters used in our model. The hyper-parameters and architectural choices were obtained by a mix of small-range random grid search and manual selection. Thus, there is little threat to the hyper-parameter choosing, and there might be room for further improvement. However, current settings have achieved a considerable performance increase.

\noindent\textbf{Threats to construct validity} relate to the suitability of our evaluation measure. We adopted the measure used by the previous method name recommendation work \citep{allamanis2016convolutional,alon2019code2vec,alon2019code2seq,nguyen2020suggesting}, which measured precision, recall, and F1 score over subtokens, and exact match accuracy. This is based on the idea that the quality of the generated method name is mostly dependant on the sub-words that were used to compose it. 

\section{Related Work}

\subsection{Code Representation}
Code representation is a hot research topic in both software engineering and machine learning fields. Different neural network-based approaches have been proposed for representing programs as vectors, which can be divided into the following categories: (1) source code token (subtoken) sequence - Using the source code token sequence as input. (2) AST node sequence - Using the flattened AST node sequence as input. (3) AST paths - Using a path through the AST as input. (4) Graph - Extending ASTs through adding edges to build the graph as input.
(5) Program entities - Using tokens in program entities’ names. These learned program vectors then can be used for various SE tasks, such as code summarization \citep{hu2018deep,wei2019code}, method name recommendation \citep{alon2019code2vec,nguyen2020suggesting}, code clone detection \citep{zhang2019novel,nafi2019clcdsa}, code completion \citep{liu2020self,liu2020multi,KarampatsisBRSJ20}, etc. These different approaches model the program from different aspects, for example, ASTs can represent the structure and the syntax of the source code better, while the graphs focus more on the data flow and the semantic of the programs. For method name recommendation, existing research mainly focuses on modeling the method body as token sequence \citep{allamanis2016convolutional,nguyen2020suggesting} or AST paths \citep{alon2019code2vec,alon2019code2seq}, and then built an RNN-based encode-decoder framework to generate the subtokens of the method name. 

\subsection{Neural Machine Translation}
Neural Machine Translation (NMT) \citep{wu2016google} is an end-to-end learning approach for automated translation. In recent years work of NMT is largely based on encoder-decoder architecture \cite{BahdanauCB14}, where the encoder maps an input sequence of words $x = (x_1, ..., x_n)$ to a sequence of continuous representations $z = (z_1, ..., z_n)$. Given $z$, the decoder then generates a sequence of output words $y = (y_1, ..., y_m)$ one token at a time, hence modeling the conditional probability: $p (y_1, ..., y_m|x_1, ..., x_n)$. The encoder-decoder architecture has been applied across many SE seq2seq tasks, including code summarization \citep{allamanis2016convolutional,hu2018deep}, method name recommendation \citep{alon2019code2vec,nguyen2020suggesting}, code generation \citep{wei2019code,sun2019grammar}, program translation \citep{chen2018tree}, etc. Different neural networks can be used in the encoder and decoder. Code2seq \cite{alon2019code2seq} employs a bi-directional LSTM to encode the AST paths then averages the representations of all the paths as the final representation of the program encoder, and employs another LSTM as the decoder to generate the output (method name or code summarization). \citet{hu2018deep} use RNN for both encoder and decoder for code comment generation task. \citet{allamanis2016convolutional} employ CNN to encode the code snippet and use GRU as decoder to generate the tokens of the method name. \citet{FernandesAB19} employ GNN as the encoder and LSTM as the decoder for a range of summarization tasks. \citet{AhmadCRC20} use transformer network for both the encoder and decoder in code summarization task.

\subsection{Method Name Recommendation}
Recommending meaningful and consistent method names is important for ensuring readability and maintainability of programs. Many approaches have been introduced to suggest succinct names for methods \citep{allamanis2016convolutional,alon2019code2vec,nguyen2020suggesting}, where different model architectures and method contexts are considered. In this section, we summarize related work on method name recommendation from the following two aspects. 

\subsubsection{Models}

\citet{suzuki2014approach} proposed an N-gram based approach to evaluate the comprehensibility of method names and suggest comprehensible method names. \citet{liu2019learning} follow an information retrieval (IR) method with the motivation that two methods with similar bodies should have similar names. They use paragraph Vector and Convolutional Neural Networks to produce the vector representations of method names and bodies, respectively. They compared the similarity of the names retrieved from the method body vector space and the method name vector space to identify the inconsistent method names. For the inconsistent names, they use the names of methods whose bodies are similar to the body of the input method to suggest the new method name. However, methods with the same bodies can still have different names since they are in different projects and are under different contexts. Besides, the IR-based approach cannot generate a new name that it has not seen before. Another kind of researches based on NMT models, where encoder-decoder framework is used to encode the method bodies and generate the method names \citep{allamanis2016convolutional,alon2019code2vec,nguyen2020suggesting}. \citet{allamanis2016convolutional} built a convolutional attentional network to extract local features of the subtoken sequence from the method body, and then use these features to suggest names for methods. \citet{alon2019code2vec} design attention-based neural network to encode the AST paths into vectors, and based on the path representation to make predictions on the method’s name. \citet{Daniel21Language} proposed Code Transformer, a Transformer-based language-agnostic code representation model. They combined distances computed on structure and context in the self-attention operation, which can learn jointly from the structure and context of programs relying on language-agnostic features. They applied their representations to the task of method name suggestion. \citet{nguyen2020suggesting} proposed an RNN-based seq2seq approach to recommend method names and to detect method name inconsistencies. They take the program entities in the method body and enclosing class name as the input. \citet{li2021context} also developed an RNN-based seq2seq approach DeepName for method name consistency checking and suggestion, which extended the contexts by considering the internal context, the caller and callee contexts, sibling context, and enclosing context. 

\subsubsection{Method Contexts}
Different method contexts are taken into account for method name recommendation. Most of the research only focused on exploiting the features from the method body, where the token sequences or ASTs of the method body are taken as the inputs. \citet{allamanis2016convolutional} considered the token sequence from the method body and built a convolutional attentional network to extract the features from the context.  \citet{alon2019code2vec}, \citet{alon2019code2seq}, \citet{Daniel21Language}, and \citet{peng2021integrating} considered the AST paths extracted from the method body as the context, and made predictions on the method’s name based on the path representation. In addition to the data from the method body, many research began to include the information from a wide range of contexts. \citet{nguyen2020suggesting} took the program entities in the method body and enclosing class name as the input. \citet{wang2021lightweight} also considered other methods in the project that have call relations with the target method. \citet{li2021context} further extended the contexts by considering the internal context, the caller and callee contexts, sibling context, and enclosing context. Inspired by these approaches, we further considered the nested scopes of the project and the documentation of the method by extracting the project-specific and documentation context, which can help for suggesting accurate method names.

\section{Conclusion}
In this paper, we propose GTNM, a global method name suggestion approach, which considers contexts of different levels, including local context, project-specific context, and the documentation of the target method. We employ a transformer-based seq2seq framework to generate the method names, which uses the attention mechanism to allow the model attending to different level contexts when generating the names. The experimental results on Java methods show that our model has a substantial improvement over baseline models.

\begin{acks}
This research is supported by the National Key R\&D Program of China under Grant No. 2020AAA0109400, and the National Natural Science Foundation of China under Grant Nos. 62072007, 62192733.
\end{acks}

\bibliographystyle{ACM-Reference-Format}
\bibliography{sample-base}


\begin{thebibliography}{48}

%%% the Web address.

\ifx \showCODEN    \undefined \def \showCODEN     #1{\unskip}     \fi
\ifx \showDOI      \undefined \def \showDOI       #1{#1}\fi
\ifx \showISBNx    \undefined \def \showISBNx     #1{\unskip}     \fi
\ifx \showISBNxiii \undefined \def \showISBNxiii  #1{\unskip}     \fi
\ifx \showISSN     \undefined \def \showISSN      #1{\unskip}     \fi
\ifx \showLCCN     \undefined \def \showLCCN      #1{\unskip}     \fi
\ifx \shownote     \undefined \def \shownote      #1{#1}          \fi
\ifx \showarticletitle \undefined \def \showarticletitle #1{#1}   \fi
\ifx \showURL      \undefined \def \showURL       {\relax}        \fi
% invisible to TeX
\providecommand\bibfield[2]{#2}
\providecommand\bibinfo[2]{#2}
\providecommand\natexlab[1]{#1}
\providecommand\showeprint[2][]{arXiv:#2}

\bibitem[\protect\citeauthoryear{Abebe, Haiduc, Tonella, and Marcus}{Abebe
  et~al\mbox{.}}{2011a}]%
        {abebe2011effect}
\bibfield{author}{\bibinfo{person}{Surafel~Lemma Abebe}, \bibinfo{person}{Sonia
  Haiduc}, \bibinfo{person}{Paolo Tonella}, {and} \bibinfo{person}{Andrian
  Marcus}.} \bibinfo{year}{2011}\natexlab{a}.
\newblock \showarticletitle{The effect of lexicon bad smells on concept
  location in source code}. In \bibinfo{booktitle}{\emph{2011 IEEE 11th
  International Working Conference on Source Code Analysis and Manipulation}}.
  Ieee, \bibinfo{pages}{125--134}.
\newblock


\bibitem[\protect\citeauthoryear{Abebe, Haiduc, Tonella, and Marcus}{Abebe
  et~al\mbox{.}}{2011b}]%
        {abebe2012can}
\bibfield{author}{\bibinfo{person}{Surafel~Lemma Abebe}, \bibinfo{person}{Sonia
  Haiduc}, \bibinfo{person}{Paolo Tonella}, {and} \bibinfo{person}{Andrian
  Marcus}.} \bibinfo{year}{2011}\natexlab{b}.
\newblock \showarticletitle{The Effect of Lexicon Bad Smells on Concept
  Location in Source Code}. In \bibinfo{booktitle}{\emph{11th {IEEE} Working
  Conference on Source Code Analysis and Manipulation, {SCAM} 2011,
  Williamsburg, VA, USA, September 25-26, 2011}}. \bibinfo{publisher}{{IEEE}
  Computer Society}, \bibinfo{pages}{125--134}.
\newblock
\urldef\tempurl%
\url{https://doi.org/10.1109/SCAM.2011.18}
\showDOI{\tempurl}


\bibitem[\protect\citeauthoryear{Ahmad, Chakraborty, Ray, and Chang}{Ahmad
  et~al\mbox{.}}{2020}]%
        {AhmadCRC20}
\bibfield{author}{\bibinfo{person}{Wasi~Uddin Ahmad}, \bibinfo{person}{Saikat
  Chakraborty}, \bibinfo{person}{Baishakhi Ray}, {and}
  \bibinfo{person}{Kai{-}Wei Chang}.} \bibinfo{year}{2020}\natexlab{}.
\newblock \showarticletitle{A Transformer-based Approach for Source Code
  Summarization}. In \bibinfo{booktitle}{\emph{Proceedings of the 58th Annual
  Meeting of the Association for Computational Linguistics, {ACL} 2020, Online,
  July 5-10, 2020}}, \bibfield{editor}{\bibinfo{person}{Dan Jurafsky},
  \bibinfo{person}{Joyce Chai}, \bibinfo{person}{Natalie Schluter}, {and}
  \bibinfo{person}{Joel~R. Tetreault}} (Eds.). \bibinfo{publisher}{Association
  for Computational Linguistics}, \bibinfo{pages}{4998--5007}.
\newblock
\urldef\tempurl%
\url{https://doi.org/10.18653/v1/2020.acl-main.449}
\showDOI{\tempurl}


\bibitem[\protect\citeauthoryear{Allamanis, Barr, Bird, and Sutton}{Allamanis
  et~al\mbox{.}}{2015}]%
        {allamanis2015suggesting}
\bibfield{author}{\bibinfo{person}{Miltiadis Allamanis},
  \bibinfo{person}{Earl~T. Barr}, \bibinfo{person}{Christian Bird}, {and}
  \bibinfo{person}{Charles Sutton}.} \bibinfo{year}{2015}\natexlab{}.
\newblock \showarticletitle{Suggesting accurate method and class names}. In
  \bibinfo{booktitle}{\emph{Proceedings of the 2015 10th Joint Meeting on
  Foundations of Software Engineering, {ESEC/FSE} 2015, Bergamo, Italy, August
  30 - September 4, 2015}}, \bibfield{editor}{\bibinfo{person}{Elisabetta~Di
  Nitto}, \bibinfo{person}{Mark Harman}, {and} \bibinfo{person}{Patrick
  Heymans}} (Eds.). \bibinfo{publisher}{{ACM}}, \bibinfo{pages}{38--49}.
\newblock
\urldef\tempurl%
\url{https://doi.org/10.1145/2786805.2786849}
\showDOI{\tempurl}


\bibitem[\protect\citeauthoryear{Allamanis, Peng, and Sutton}{Allamanis
  et~al\mbox{.}}{2016}]%
        {allamanis2016convolutional}
\bibfield{author}{\bibinfo{person}{Miltiadis Allamanis}, \bibinfo{person}{Hao
  Peng}, {and} \bibinfo{person}{Charles Sutton}.}
  \bibinfo{year}{2016}\natexlab{}.
\newblock \showarticletitle{A Convolutional Attention Network for Extreme
  Summarization of Source Code}. In \bibinfo{booktitle}{\emph{Proceedings of
  the 33nd International Conference on Machine Learning, {ICML} 2016, New York
  City, NY, USA, June 19-24, 2016}} \emph{(\bibinfo{series}{{JMLR} Workshop and
  Conference Proceedings}, Vol.~\bibinfo{volume}{48})},
  \bibfield{editor}{\bibinfo{person}{Maria{-}Florina Balcan} {and}
  \bibinfo{person}{Kilian~Q. Weinberger}} (Eds.).
  \bibinfo{publisher}{JMLR.org}, \bibinfo{pages}{2091--2100}.
\newblock
\urldef\tempurl%
\url{http://proceedings.mlr.press/v48/allamanis16.html}
\showURL{%
\tempurl}


\bibitem[\protect\citeauthoryear{Alon, Brody, Levy, and Yahav}{Alon
  et~al\mbox{.}}{2019a}]%
        {alon2019code2seq}
\bibfield{author}{\bibinfo{person}{Uri Alon}, \bibinfo{person}{Shaked Brody},
  \bibinfo{person}{Omer Levy}, {and} \bibinfo{person}{Eran Yahav}.}
  \bibinfo{year}{2019}\natexlab{a}.
\newblock \showarticletitle{code2seq: Generating Sequences from Structured
  Representations of Code}. In \bibinfo{booktitle}{\emph{7th International
  Conference on Learning Representations, {ICLR} 2019, New Orleans, LA, USA,
  May 6-9, 2019}}. \bibinfo{publisher}{OpenReview.net}.
\newblock
\urldef\tempurl%
\url{https://openreview.net/forum?id=H1gKYo09tX}
\showURL{%
\tempurl}


\bibitem[\protect\citeauthoryear{Alon, Zilberstein, Levy, and Yahav}{Alon
  et~al\mbox{.}}{2019b}]%
        {alon2019code2vec}
\bibfield{author}{\bibinfo{person}{Uri Alon}, \bibinfo{person}{Meital
  Zilberstein}, \bibinfo{person}{Omer Levy}, {and} \bibinfo{person}{Eran
  Yahav}.} \bibinfo{year}{2019}\natexlab{b}.
\newblock \showarticletitle{code2vec: learning distributed representations of
  code}.
\newblock \bibinfo{journal}{\emph{Proc. {ACM} Program. Lang.}}
  \bibinfo{volume}{3}, \bibinfo{number}{{POPL}} (\bibinfo{year}{2019}),
  \bibinfo{pages}{40:1--40:29}.
\newblock
\urldef\tempurl%
\url{https://doi.org/10.1145/3290353}
\showDOI{\tempurl}


\bibitem[\protect\citeauthoryear{Amann, Nguyen, Nadi, Nguyen, and Mezini}{Amann
  et~al\mbox{.}}{2019}]%
        {amann2018systematic}
\bibfield{author}{\bibinfo{person}{Sven Amann}, \bibinfo{person}{Hoan~Anh
  Nguyen}, \bibinfo{person}{Sarah Nadi}, \bibinfo{person}{Tien~N. Nguyen},
  {and} \bibinfo{person}{Mira Mezini}.} \bibinfo{year}{2019}\natexlab{}.
\newblock \showarticletitle{A Systematic Evaluation of Static API-Misuse
  Detectors}.
\newblock \bibinfo{journal}{\emph{{IEEE} Trans. Software Eng.}}
  \bibinfo{volume}{45}, \bibinfo{number}{12} (\bibinfo{year}{2019}),
  \bibinfo{pages}{1170--1188}.
\newblock
\urldef\tempurl%
\url{https://doi.org/10.1109/TSE.2018.2827384}
\showDOI{\tempurl}


\bibitem[\protect\citeauthoryear{Arnaoudova, Eshkevari, Penta, Oliveto,
  Antoniol, and Gu{\'{e}}h{\'{e}}neuc}{Arnaoudova et~al\mbox{.}}{2014}]%
        {arnaoudova2014repent}
\bibfield{author}{\bibinfo{person}{Venera Arnaoudova},
  \bibinfo{person}{Laleh~Mousavi Eshkevari}, \bibinfo{person}{Massimiliano~Di
  Penta}, \bibinfo{person}{Rocco Oliveto}, \bibinfo{person}{Giuliano Antoniol},
  {and} \bibinfo{person}{Yann{-}Ga{\"{e}}l Gu{\'{e}}h{\'{e}}neuc}.}
  \bibinfo{year}{2014}\natexlab{}.
\newblock \showarticletitle{{REPENT:} Analyzing the Nature of Identifier
  Renamings}.
\newblock \bibinfo{journal}{\emph{{IEEE} Trans. Software Eng.}}
  \bibinfo{volume}{40}, \bibinfo{number}{5} (\bibinfo{year}{2014}),
  \bibinfo{pages}{502--532}.
\newblock
\urldef\tempurl%
\url{https://doi.org/10.1109/TSE.2014.2312942}
\showDOI{\tempurl}


\bibitem[\protect\citeauthoryear{Arnaoudova, Penta, and Antoniol}{Arnaoudova
  et~al\mbox{.}}{2016}]%
        {arnaoudova2016linguistic}
\bibfield{author}{\bibinfo{person}{Venera Arnaoudova},
  \bibinfo{person}{Massimiliano~Di Penta}, {and} \bibinfo{person}{Giuliano
  Antoniol}.} \bibinfo{year}{2016}\natexlab{}.
\newblock \showarticletitle{Linguistic antipatterns: what they are and how
  developers perceive them}.
\newblock \bibinfo{journal}{\emph{Empir. Softw. Eng.}} \bibinfo{volume}{21},
  \bibinfo{number}{1} (\bibinfo{year}{2016}), \bibinfo{pages}{104--158}.
\newblock
\urldef\tempurl%
\url{https://doi.org/10.1007/s10664-014-9350-8}
\showDOI{\tempurl}


\bibitem[\protect\citeauthoryear{Bahdanau, Cho, and Bengio}{Bahdanau
  et~al\mbox{.}}{2015}]%
        {BahdanauCB14}
\bibfield{author}{\bibinfo{person}{Dzmitry Bahdanau},
  \bibinfo{person}{Kyunghyun Cho}, {and} \bibinfo{person}{Yoshua Bengio}.}
  \bibinfo{year}{2015}\natexlab{}.
\newblock \showarticletitle{Neural Machine Translation by Jointly Learning to
  Align and Translate}. In \bibinfo{booktitle}{\emph{3rd International
  Conference on Learning Representations, {ICLR} 2015, San Diego, CA, USA, May
  7-9, 2015, Conference Track Proceedings}},
  \bibfield{editor}{\bibinfo{person}{Yoshua Bengio} {and} \bibinfo{person}{Yann
  LeCun}} (Eds.).
\newblock
\urldef\tempurl%
\url{http://arxiv.org/abs/1409.0473}
\showURL{%
\tempurl}


\bibitem[\protect\citeauthoryear{Beck}{Beck}{2007}]%
        {beck2007implementation}
\bibfield{author}{\bibinfo{person}{Kent Beck}.}
  \bibinfo{year}{2007}\natexlab{}.
\newblock \bibinfo{booktitle}{\emph{Implementation patterns}}.
\newblock \bibinfo{publisher}{Pearson Education}.
\newblock


\bibitem[\protect\citeauthoryear{Butler, Wermelinger, Yu, and Sharp}{Butler
  et~al\mbox{.}}{2009}]%
        {butler2009relating}
\bibfield{author}{\bibinfo{person}{Simon Butler}, \bibinfo{person}{Michel
  Wermelinger}, \bibinfo{person}{Yijun Yu}, {and} \bibinfo{person}{Helen
  Sharp}.} \bibinfo{year}{2009}\natexlab{}.
\newblock \showarticletitle{Relating Identifier Naming Flaws and Code Quality:
  An Empirical Study}. In \bibinfo{booktitle}{\emph{16th Working Conference on
  Reverse Engineering, {WCRE} 2009, 13-16 October 2009, Lille, France}},
  \bibfield{editor}{\bibinfo{person}{Andy Zaidman}, \bibinfo{person}{Giuliano
  Antoniol}, {and} \bibinfo{person}{St{\'{e}}phane Ducasse}} (Eds.).
  \bibinfo{publisher}{{IEEE} Computer Society}, \bibinfo{pages}{31--35}.
\newblock
\urldef\tempurl%
\url{https://doi.org/10.1109/WCRE.2009.50}
\showDOI{\tempurl}


\bibitem[\protect\citeauthoryear{Chen, Liu, and Song}{Chen
  et~al\mbox{.}}{2018}]%
        {chen2018tree}
\bibfield{author}{\bibinfo{person}{Xinyun Chen}, \bibinfo{person}{Chang Liu},
  {and} \bibinfo{person}{Dawn Song}.} \bibinfo{year}{2018}\natexlab{}.
\newblock \showarticletitle{Tree-to-tree Neural Networks for Program
  Translation}. In \bibinfo{booktitle}{\emph{Advances in Neural Information
  Processing Systems 31: Annual Conference on Neural Information Processing
  Systems 2018, NeurIPS 2018, December 3-8, 2018, Montr{\'{e}}al, Canada}},
  \bibfield{editor}{\bibinfo{person}{Samy Bengio}, \bibinfo{person}{Hanna~M.
  Wallach}, \bibinfo{person}{Hugo Larochelle}, \bibinfo{person}{Kristen
  Grauman}, \bibinfo{person}{Nicol{\`{o}} Cesa{-}Bianchi}, {and}
  \bibinfo{person}{Roman Garnett}} (Eds.). \bibinfo{pages}{2552--2562}.
\newblock
\urldef\tempurl%
\url{https://proceedings.neurips.cc/paper/2018/hash/d759175de8ea5b1d9a2660e45554894f-Abstract.html}
\showURL{%
\tempurl}


\bibitem[\protect\citeauthoryear{Fernandes, Allamanis, and
  Brockschmidt}{Fernandes et~al\mbox{.}}{2019}]%
        {FernandesAB19}
\bibfield{author}{\bibinfo{person}{Patrick Fernandes},
  \bibinfo{person}{Miltiadis Allamanis}, {and} \bibinfo{person}{Marc
  Brockschmidt}.} \bibinfo{year}{2019}\natexlab{}.
\newblock \showarticletitle{Structured Neural Summarization}. In
  \bibinfo{booktitle}{\emph{7th International Conference on Learning
  Representations, {ICLR} 2019, New Orleans, LA, USA, May 6-9, 2019}}.
  \bibinfo{publisher}{OpenReview.net}.
\newblock
\urldef\tempurl%
\url{https://openreview.net/forum?id=H1ersoRqtm}
\showURL{%
\tempurl}


\bibitem[\protect\citeauthoryear{Hindle, Barr, Gabel, Su, and Devanbu}{Hindle
  et~al\mbox{.}}{2016}]%
        {hindle2016naturalness}
\bibfield{author}{\bibinfo{person}{Abram Hindle}, \bibinfo{person}{Earl~T
  Barr}, \bibinfo{person}{Mark Gabel}, \bibinfo{person}{Zhendong Su}, {and}
  \bibinfo{person}{Premkumar Devanbu}.} \bibinfo{year}{2016}\natexlab{}.
\newblock \showarticletitle{On the naturalness of software}.
\newblock \bibinfo{journal}{\emph{Commun. ACM}} \bibinfo{volume}{59},
  \bibinfo{number}{5} (\bibinfo{year}{2016}), \bibinfo{pages}{122--131}.
\newblock


\bibitem[\protect\citeauthoryear{Hofmeister, Siegmund, and Holt}{Hofmeister
  et~al\mbox{.}}{2017}]%
        {hofmeister2017shorter}
\bibfield{author}{\bibinfo{person}{Johannes~C. Hofmeister},
  \bibinfo{person}{Janet Siegmund}, {and} \bibinfo{person}{Daniel~V. Holt}.}
  \bibinfo{year}{2017}\natexlab{}.
\newblock \showarticletitle{Shorter identifier names take longer to
  comprehend}. In \bibinfo{booktitle}{\emph{{IEEE} 24th International
  Conference on Software Analysis, Evolution and Reengineering, {SANER} 2017,
  Klagenfurt, Austria, February 20-24, 2017}},
  \bibfield{editor}{\bibinfo{person}{Martin Pinzger}, \bibinfo{person}{Gabriele
  Bavota}, {and} \bibinfo{person}{Andrian Marcus}} (Eds.).
  \bibinfo{publisher}{{IEEE} Computer Society}, \bibinfo{pages}{217--227}.
\newblock
\urldef\tempurl%
\url{https://doi.org/10.1109/SANER.2017.7884623}
\showDOI{\tempurl}


\bibitem[\protect\citeauthoryear{H{\o}st and {\O}stvold}{H{\o}st and
  {\O}stvold}{2009}]%
        {host2009debugging}
\bibfield{author}{\bibinfo{person}{Einar~W. H{\o}st} {and}
  \bibinfo{person}{Bjarte~M. {\O}stvold}.} \bibinfo{year}{2009}\natexlab{}.
\newblock \showarticletitle{Debugging Method Names}. In
  \bibinfo{booktitle}{\emph{{ECOOP} 2009 - Object-Oriented Programming, 23rd
  European Conference, Genoa, Italy, July 6-10, 2009. Proceedings}}
  \emph{(\bibinfo{series}{Lecture Notes in Computer Science},
  Vol.~\bibinfo{volume}{5653})}, \bibfield{editor}{\bibinfo{person}{Sophia
  Drossopoulou}} (Ed.). \bibinfo{publisher}{Springer},
  \bibinfo{pages}{294--317}.
\newblock
\urldef\tempurl%
\url{https://doi.org/10.1007/978-3-642-03013-0\_14}
\showDOI{\tempurl}


\bibitem[\protect\citeauthoryear{Hu, Li, Xia, Lo, and Jin}{Hu
  et~al\mbox{.}}{2018}]%
        {hu2018deep}
\bibfield{author}{\bibinfo{person}{Xing Hu}, \bibinfo{person}{Ge Li},
  \bibinfo{person}{Xin Xia}, \bibinfo{person}{David Lo}, {and}
  \bibinfo{person}{Zhi Jin}.} \bibinfo{year}{2018}\natexlab{}.
\newblock \showarticletitle{Deep code comment generation}. In
  \bibinfo{booktitle}{\emph{Proceedings of the 26th Conference on Program
  Comprehension, {ICPC} 2018, Gothenburg, Sweden, May 27-28, 2018}},
  \bibfield{editor}{\bibinfo{person}{Foutse Khomh},
  \bibinfo{person}{Chanchal~K. Roy}, {and} \bibinfo{person}{Janet Siegmund}}
  (Eds.). \bibinfo{publisher}{{ACM}}, \bibinfo{pages}{200--210}.
\newblock
\urldef\tempurl%
\url{https://doi.org/10.1145/3196321.3196334}
\showDOI{\tempurl}


\bibitem[\protect\citeauthoryear{Hu, Li, Xia, Lo, and Jin}{Hu
  et~al\mbox{.}}{2020}]%
        {hu2020deep}
\bibfield{author}{\bibinfo{person}{Xing Hu}, \bibinfo{person}{Ge Li},
  \bibinfo{person}{Xin Xia}, \bibinfo{person}{David Lo}, {and}
  \bibinfo{person}{Zhi Jin}.} \bibinfo{year}{2020}\natexlab{}.
\newblock \showarticletitle{Deep code comment generation with hybrid lexical
  and syntactical information}.
\newblock \bibinfo{journal}{\emph{Empirical Software Engineering}}
  \bibinfo{volume}{25}, \bibinfo{number}{3} (\bibinfo{year}{2020}),
  \bibinfo{pages}{2179--2217}.
\newblock


\bibitem[\protect\citeauthoryear{Karampatsis, Babii, Robbes, Sutton, and
  Janes}{Karampatsis et~al\mbox{.}}{2020}]%
        {KarampatsisBRSJ20}
\bibfield{author}{\bibinfo{person}{Rafael{-}Michael Karampatsis},
  \bibinfo{person}{Hlib Babii}, \bibinfo{person}{Romain Robbes},
  \bibinfo{person}{Charles Sutton}, {and} \bibinfo{person}{Andrea Janes}.}
  \bibinfo{year}{2020}\natexlab{}.
\newblock \showarticletitle{Big code != big vocabulary: open-vocabulary models
  for source code}. In \bibinfo{booktitle}{\emph{{ICSE} '20: 42nd International
  Conference on Software Engineering, Seoul, South Korea, 27 June - 19 July,
  2020}}, \bibfield{editor}{\bibinfo{person}{Gregg Rothermel} {and}
  \bibinfo{person}{Doo{-}Hwan Bae}} (Eds.). \bibinfo{publisher}{{ACM}},
  \bibinfo{pages}{1073--1085}.
\newblock
\urldef\tempurl%
\url{https://doi.org/10.1145/3377811.3380342}
\showDOI{\tempurl}


\bibitem[\protect\citeauthoryear{Lawrie, Morrell, Feild, and Binkley}{Lawrie
  et~al\mbox{.}}{2006}]%
        {lawrie2006s}
\bibfield{author}{\bibinfo{person}{Dawn~J. Lawrie},
  \bibinfo{person}{Christopher Morrell}, \bibinfo{person}{Henry Feild}, {and}
  \bibinfo{person}{David~W. Binkley}.} \bibinfo{year}{2006}\natexlab{}.
\newblock \showarticletitle{What's in a Name? {A} Study of Identifiers}. In
  \bibinfo{booktitle}{\emph{14th International Conference on Program
  Comprehension {(ICPC} 2006), 14-16 June 2006, Athens, Greece}}.
  \bibinfo{publisher}{{IEEE} Computer Society}, \bibinfo{pages}{3--12}.
\newblock
\urldef\tempurl%
\url{https://doi.org/10.1109/ICPC.2006.51}
\showDOI{\tempurl}


\bibitem[\protect\citeauthoryear{LeClair, Jiang, and McMillan}{LeClair
  et~al\mbox{.}}{2019}]%
        {leclair2019neural}
\bibfield{author}{\bibinfo{person}{Alexander LeClair}, \bibinfo{person}{Siyuan
  Jiang}, {and} \bibinfo{person}{Collin McMillan}.}
  \bibinfo{year}{2019}\natexlab{}.
\newblock \showarticletitle{A neural model for generating natural language
  summaries of program subroutines}. In \bibinfo{booktitle}{\emph{2019 IEEE/ACM
  41st International Conference on Software Engineering (ICSE)}}. IEEE,
  \bibinfo{pages}{795--806}.
\newblock


\bibitem[\protect\citeauthoryear{Li, Wang, and Nguyen}{Li
  et~al\mbox{.}}{2021}]%
        {li2021context}
\bibfield{author}{\bibinfo{person}{Yi Li}, \bibinfo{person}{Shaohua Wang},
  {and} \bibinfo{person}{Tien~N Nguyen}.} \bibinfo{year}{2021}\natexlab{}.
\newblock \showarticletitle{A Context-based Automated Approach for Method Name
  Consistency Checking and Suggestion}. In \bibinfo{booktitle}{\emph{2021
  IEEE/ACM 43rd International Conference on Software Engineering (ICSE)}}.
  IEEE, \bibinfo{pages}{574--586}.
\newblock


\bibitem[\protect\citeauthoryear{Liblit, Begel, and Sweetser}{Liblit
  et~al\mbox{.}}{2006}]%
        {liblit2006cognitive}
\bibfield{author}{\bibinfo{person}{Ben Liblit}, \bibinfo{person}{Andrew Begel},
  {and} \bibinfo{person}{Eve Sweetser}.} \bibinfo{year}{2006}\natexlab{}.
\newblock \showarticletitle{Cognitive Perspectives on the Role of Naming in
  Computer Programs}. In \bibinfo{booktitle}{\emph{Proceedings of the 18th
  Annual Workshop of the Psychology of Programming Interest Group, {PPIG} 2006,
  Brighton, UK, September 7-8, 2006}}. \bibinfo{publisher}{Psychology of
  Programming Interest Group}, \bibinfo{pages}{11}.
\newblock
\urldef\tempurl%
\url{http://ppig.org/library/paper/cognitive-perspectives-role-naming-computer-programs}
\showURL{%
\tempurl}


\bibitem[\protect\citeauthoryear{Liu, Li, Wei, Xia, Fu, and Jin}{Liu
  et~al\mbox{.}}{2020a}]%
        {liu2020self}
\bibfield{author}{\bibinfo{person}{Fang Liu}, \bibinfo{person}{Ge Li},
  \bibinfo{person}{Bolin Wei}, \bibinfo{person}{Xin Xia},
  \bibinfo{person}{Zhiyi Fu}, {and} \bibinfo{person}{Zhi Jin}.}
  \bibinfo{year}{2020}\natexlab{a}.
\newblock \showarticletitle{A Self-Attentional Neural Architecture for Code
  Completion with Multi-Task Learning}. In \bibinfo{booktitle}{\emph{{ICPC}
  '20: 28th International Conference on Program Comprehension, Seoul, Republic
  of Korea, July 13-15, 2020}}. \bibinfo{publisher}{{ACM}},
  \bibinfo{pages}{37--47}.
\newblock
\urldef\tempurl%
\url{https://doi.org/10.1145/3387904.3389261}
\showDOI{\tempurl}


\bibitem[\protect\citeauthoryear{Liu, Li, Zhao, and Jin}{Liu
  et~al\mbox{.}}{2020b}]%
        {liu2020multi}
\bibfield{author}{\bibinfo{person}{Fang Liu}, \bibinfo{person}{Ge Li},
  \bibinfo{person}{Yunfei Zhao}, {and} \bibinfo{person}{Zhi Jin}.}
  \bibinfo{year}{2020}\natexlab{b}.
\newblock \showarticletitle{Multi-task Learning based Pre-trained Language
  Model for Code Completion}. In \bibinfo{booktitle}{\emph{35th {IEEE/ACM}
  International Conference on Automated Software Engineering, {ASE} 2020,
  Melbourne, Australia, September 21-25, 2020}}. \bibinfo{publisher}{{IEEE}},
  \bibinfo{pages}{473--485}.
\newblock
\urldef\tempurl%
\url{https://doi.org/10.1145/3324884.3416591}
\showDOI{\tempurl}


\bibitem[\protect\citeauthoryear{Liu, Kim, Bissyand{\'{e}}, Kim, Kim, Koyuncu,
  Kim, and Traon}{Liu et~al\mbox{.}}{2019}]%
        {liu2019learning}
\bibfield{author}{\bibinfo{person}{Kui Liu}, \bibinfo{person}{Dongsun Kim},
  \bibinfo{person}{Tegawend{\'{e}}~F. Bissyand{\'{e}}},
  \bibinfo{person}{Tae{-}young Kim}, \bibinfo{person}{Kisub Kim},
  \bibinfo{person}{Anil Koyuncu}, \bibinfo{person}{Suntae Kim}, {and}
  \bibinfo{person}{Yves~Le Traon}.} \bibinfo{year}{2019}\natexlab{}.
\newblock \showarticletitle{Learning to spot and refactor inconsistent method
  names}. In \bibinfo{booktitle}{\emph{Proceedings of the 41st International
  Conference on Software Engineering, {ICSE} 2019, Montreal, QC, Canada, May
  25-31, 2019}}, \bibfield{editor}{\bibinfo{person}{Joanne~M. Atlee},
  \bibinfo{person}{Tevfik Bultan}, {and} \bibinfo{person}{Jon Whittle}} (Eds.).
  \bibinfo{publisher}{{IEEE} / {ACM}}, \bibinfo{pages}{1--12}.
\newblock
\urldef\tempurl%
\url{https://doi.org/10.1109/ICSE.2019.00019}
\showDOI{\tempurl}


\bibitem[\protect\citeauthoryear{Macbeth, Razumiejczyk, and Ledesma}{Macbeth
  et~al\mbox{.}}{2011}]%
        {macbeth2011cliff}
\bibfield{author}{\bibinfo{person}{Guillermo Macbeth}, \bibinfo{person}{Eugenia
  Razumiejczyk}, {and} \bibinfo{person}{Rub{\'e}n~Daniel Ledesma}.}
  \bibinfo{year}{2011}\natexlab{}.
\newblock \showarticletitle{Cliff's Delta Calculator: A non-parametric effect
  size program for two groups of observations}.
\newblock \bibinfo{journal}{\emph{Universitas Psychologica}}
  \bibinfo{volume}{10}, \bibinfo{number}{2} (\bibinfo{year}{2011}),
  \bibinfo{pages}{545--555}.
\newblock


\bibitem[\protect\citeauthoryear{Martin}{Martin}{2009}]%
        {martin2009clean}
\bibfield{author}{\bibinfo{person}{Robert~C. Martin}.}
  \bibinfo{year}{2009}\natexlab{}.
\newblock \bibinfo{booktitle}{\emph{Clean Code - a Handbook of Agile Software
  Craftsmanship}}.
\newblock \bibinfo{publisher}{Prentice Hall}.
\newblock
\showISBNx{978-0-13-235088-4}
\urldef\tempurl%
\url{http://vig.pearsoned.com/store/product/1,1207,store-12521\_isbn-0132350882,00.html}
\showURL{%
\tempurl}


\bibitem[\protect\citeauthoryear{McBurney and McMillan}{McBurney and
  McMillan}{2015}]%
        {mcburney2015automatic}
\bibfield{author}{\bibinfo{person}{Paul~W McBurney} {and}
  \bibinfo{person}{Collin McMillan}.} \bibinfo{year}{2015}\natexlab{}.
\newblock \showarticletitle{Automatic source code summarization of context for
  java methods}.
\newblock \bibinfo{journal}{\emph{IEEE Transactions on Software Engineering}}
  \bibinfo{volume}{42}, \bibinfo{number}{2} (\bibinfo{year}{2015}),
  \bibinfo{pages}{103--119}.
\newblock


\bibitem[\protect\citeauthoryear{McConnell}{McConnell}{2004}]%
        {mcconnell2004code}
\bibfield{author}{\bibinfo{person}{Steve McConnell}.}
  \bibinfo{year}{2004}\natexlab{}.
\newblock \bibinfo{booktitle}{\emph{Code complete - a practical handbook of
  software construction, 2nd Edition}}.
\newblock \bibinfo{publisher}{Microsoft Press}.
\newblock
\showISBNx{9780735619678}
\urldef\tempurl%
\url{https://www.worldcat.org/oclc/249645389}
\showURL{%
\tempurl}


\bibitem[\protect\citeauthoryear{Nafi, Kar, Roy, Roy, and Schneider}{Nafi
  et~al\mbox{.}}{2019}]%
        {nafi2019clcdsa}
\bibfield{author}{\bibinfo{person}{Kawser~Wazed Nafi},
  \bibinfo{person}{Tonny~Shekha Kar}, \bibinfo{person}{Banani Roy},
  \bibinfo{person}{Chanchal~K. Roy}, {and} \bibinfo{person}{Kevin~A.
  Schneider}.} \bibinfo{year}{2019}\natexlab{}.
\newblock \showarticletitle{{CLCDSA:} Cross Language Code Clone Detection using
  Syntactical Features and {API} Documentation}. In
  \bibinfo{booktitle}{\emph{34th {IEEE/ACM} International Conference on
  Automated Software Engineering, {ASE} 2019, San Diego, CA, USA, November
  11-15, 2019}}. \bibinfo{publisher}{{IEEE}}, \bibinfo{pages}{1026--1037}.
\newblock
\urldef\tempurl%
\url{https://doi.org/10.1109/ASE.2019.00099}
\showDOI{\tempurl}


\bibitem[\protect\citeauthoryear{Nguyen, Phan, Le, and Nguyen}{Nguyen
  et~al\mbox{.}}{2020}]%
        {nguyen2020suggesting}
\bibfield{author}{\bibinfo{person}{Son Nguyen}, \bibinfo{person}{Hung Phan},
  \bibinfo{person}{Trinh Le}, {and} \bibinfo{person}{Tien~N. Nguyen}.}
  \bibinfo{year}{2020}\natexlab{}.
\newblock \showarticletitle{Suggesting natural method names to check name
  consistencies}. In \bibinfo{booktitle}{\emph{{ICSE} '20: 42nd International
  Conference on Software Engineering, Seoul, South Korea, 27 June - 19 July,
  2020}}, \bibfield{editor}{\bibinfo{person}{Gregg Rothermel} {and}
  \bibinfo{person}{Doo{-}Hwan Bae}} (Eds.). \bibinfo{publisher}{{ACM}},
  \bibinfo{pages}{1372--1384}.
\newblock
\urldef\tempurl%
\url{https://doi.org/10.1145/3377811.3380926}
\showDOI{\tempurl}


\bibitem[\protect\citeauthoryear{Pei, Cao, Yang, and Jana}{Pei
  et~al\mbox{.}}{2017}]%
        {pei2017deepxplore}
\bibfield{author}{\bibinfo{person}{Kexin Pei}, \bibinfo{person}{Yinzhi Cao},
  \bibinfo{person}{Junfeng Yang}, {and} \bibinfo{person}{Suman Jana}.}
  \bibinfo{year}{2017}\natexlab{}.
\newblock \showarticletitle{Deepxplore: Automated whitebox testing of deep
  learning systems}. In \bibinfo{booktitle}{\emph{proceedings of the 26th
  Symposium on Operating Systems Principles}}. \bibinfo{pages}{1--18}.
\newblock


\bibitem[\protect\citeauthoryear{Peng, Li, Wang, Zhao, and Jin}{Peng
  et~al\mbox{.}}{2021}]%
        {peng2021integrating}
\bibfield{author}{\bibinfo{person}{Han Peng}, \bibinfo{person}{Ge Li},
  \bibinfo{person}{Wenhan Wang}, \bibinfo{person}{Yunfei Zhao}, {and}
  \bibinfo{person}{Zhi Jin}.} \bibinfo{year}{2021}\natexlab{}.
\newblock \showarticletitle{Integrating Tree Path in Transformer for Code
  Representation}.
\newblock \bibinfo{journal}{\emph{Advances in Neural Information Processing
  Systems}}  \bibinfo{volume}{34} (\bibinfo{year}{2021}).
\newblock


\bibitem[\protect\citeauthoryear{Sun, Zhu, Mou, Xiong, Li, and Zhang}{Sun
  et~al\mbox{.}}{2019}]%
        {sun2019grammar}
\bibfield{author}{\bibinfo{person}{Zeyu Sun}, \bibinfo{person}{Qihao Zhu},
  \bibinfo{person}{Lili Mou}, \bibinfo{person}{Yingfei Xiong},
  \bibinfo{person}{Ge Li}, {and} \bibinfo{person}{Lu Zhang}.}
  \bibinfo{year}{2019}\natexlab{}.
\newblock \showarticletitle{A Grammar-Based Structural {CNN} Decoder for Code
  Generation}. In \bibinfo{booktitle}{\emph{The Thirty-Third {AAAI} Conference
  on Artificial Intelligence, {AAAI} 2019, The Thirty-First Innovative
  Applications of Artificial Intelligence Conference, {IAAI} 2019, The Ninth
  {AAAI} Symposium on Educational Advances in Artificial Intelligence, {EAAI}
  2019, Honolulu, Hawaii, USA, January 27 - February 1, 2019}}.
  \bibinfo{publisher}{{AAAI} Press}, \bibinfo{pages}{7055--7062}.
\newblock
\urldef\tempurl%
\url{https://doi.org/10.1609/aaai.v33i01.33017055}
\showDOI{\tempurl}


\bibitem[\protect\citeauthoryear{Suzuki, Sakamoto, Ishikawa, and
  Honiden}{Suzuki et~al\mbox{.}}{2014}]%
        {suzuki2014approach}
\bibfield{author}{\bibinfo{person}{Takayuki Suzuki}, \bibinfo{person}{Kazunori
  Sakamoto}, \bibinfo{person}{Fuyuki Ishikawa}, {and} \bibinfo{person}{Shinichi
  Honiden}.} \bibinfo{year}{2014}\natexlab{}.
\newblock \showarticletitle{An approach for evaluating and suggesting method
  names using n-gram models}. In \bibinfo{booktitle}{\emph{22nd International
  Conference on Program Comprehension, {ICPC} 2014, Hyderabad, India, June 2-3,
  2014}}, \bibfield{editor}{\bibinfo{person}{Chanchal~K. Roy},
  \bibinfo{person}{Andrew Begel}, {and} \bibinfo{person}{Leon Moonen}} (Eds.).
  \bibinfo{publisher}{{ACM}}, \bibinfo{pages}{271--274}.
\newblock
\urldef\tempurl%
\url{https://doi.org/10.1145/2597008.2597797}
\showDOI{\tempurl}


\bibitem[\protect\citeauthoryear{Takang, Grubb, and Macredie}{Takang
  et~al\mbox{.}}{1996}]%
        {takang1996effects}
\bibfield{author}{\bibinfo{person}{Armstrong~A. Takang},
  \bibinfo{person}{Penny~A. Grubb}, {and} \bibinfo{person}{Robert~D.
  Macredie}.} \bibinfo{year}{1996}\natexlab{}.
\newblock \showarticletitle{The effects of comments and identifier names on
  program comprehensibility: an experimental investigation}.
\newblock \bibinfo{journal}{\emph{J. Program. Lang.}} \bibinfo{volume}{4},
  \bibinfo{number}{3} (\bibinfo{year}{1996}), \bibinfo{pages}{143--167}.
\newblock
\urldef\tempurl%
\url{http://compscinet.dcs.kcl.ac.uk/JP/jp040302.abs.html}
\showURL{%
\tempurl}


\bibitem[\protect\citeauthoryear{Tian, Pei, Jana, and Ray}{Tian
  et~al\mbox{.}}{2018}]%
        {tian2018deeptest}
\bibfield{author}{\bibinfo{person}{Yuchi Tian}, \bibinfo{person}{Kexin Pei},
  \bibinfo{person}{Suman Jana}, {and} \bibinfo{person}{Baishakhi Ray}.}
  \bibinfo{year}{2018}\natexlab{}.
\newblock \showarticletitle{Deeptest: Automated testing of
  deep-neural-network-driven autonomous cars}. In
  \bibinfo{booktitle}{\emph{Proceedings of the 40th international conference on
  software engineering}}. \bibinfo{pages}{303--314}.
\newblock


\bibitem[\protect\citeauthoryear{Vaswani, Shazeer, Parmar, Uszkoreit, Jones,
  Gomez, Kaiser, and Polosukhin}{Vaswani et~al\mbox{.}}{2017}]%
        {vaswani2017attention}
\bibfield{author}{\bibinfo{person}{Ashish Vaswani}, \bibinfo{person}{Noam
  Shazeer}, \bibinfo{person}{Niki Parmar}, \bibinfo{person}{Jakob Uszkoreit},
  \bibinfo{person}{Llion Jones}, \bibinfo{person}{Aidan~N. Gomez},
  \bibinfo{person}{Lukasz Kaiser}, {and} \bibinfo{person}{Illia Polosukhin}.}
  \bibinfo{year}{2017}\natexlab{}.
\newblock \showarticletitle{Attention is All you Need}. In
  \bibinfo{booktitle}{\emph{Advances in Neural Information Processing Systems
  30: Annual Conference on Neural Information Processing Systems 2017, December
  4-9, 2017, Long Beach, CA, {USA}}},
  \bibfield{editor}{\bibinfo{person}{Isabelle Guyon}, \bibinfo{person}{Ulrike
  von Luxburg}, \bibinfo{person}{Samy Bengio}, \bibinfo{person}{Hanna~M.
  Wallach}, \bibinfo{person}{Rob Fergus}, \bibinfo{person}{S.~V.~N.
  Vishwanathan}, {and} \bibinfo{person}{Roman Garnett}} (Eds.).
  \bibinfo{pages}{5998--6008}.
\newblock
\urldef\tempurl%
\url{https://proceedings.neurips.cc/paper/2017/hash/3f5ee243547dee91fbd053c1c4a845aa-Abstract.html}
\showURL{%
\tempurl}


\bibitem[\protect\citeauthoryear{Wang, Wen, Lin, and Mao}{Wang
  et~al\mbox{.}}{2021}]%
        {wang2021lightweight}
\bibfield{author}{\bibinfo{person}{Shangwen Wang}, \bibinfo{person}{Ming Wen},
  \bibinfo{person}{Bo Lin}, {and} \bibinfo{person}{Xiaoguang Mao}.}
  \bibinfo{year}{2021}\natexlab{}.
\newblock \showarticletitle{Lightweight global and local contexts guided method
  name recommendation with prior knowledge}. In
  \bibinfo{booktitle}{\emph{Proceedings of the 29th ACM Joint Meeting on
  European Software Engineering Conference and Symposium on the Foundations of
  Software Engineering}}. \bibinfo{pages}{741--753}.
\newblock


\bibitem[\protect\citeauthoryear{Wei, Li, Xia, Fu, and Jin}{Wei
  et~al\mbox{.}}{2019}]%
        {wei2019code}
\bibfield{author}{\bibinfo{person}{Bolin Wei}, \bibinfo{person}{Ge Li},
  \bibinfo{person}{Xin Xia}, \bibinfo{person}{Zhiyi Fu}, {and}
  \bibinfo{person}{Zhi Jin}.} \bibinfo{year}{2019}\natexlab{}.
\newblock \showarticletitle{Code Generation as a Dual Task of Code
  Summarization}. In \bibinfo{booktitle}{\emph{Advances in Neural Information
  Processing Systems 32: Annual Conference on Neural Information Processing
  Systems 2019, NeurIPS 2019, December 8-14, 2019, Vancouver, BC, Canada}},
  \bibfield{editor}{\bibinfo{person}{Hanna~M. Wallach}, \bibinfo{person}{Hugo
  Larochelle}, \bibinfo{person}{Alina Beygelzimer}, \bibinfo{person}{Florence
  d'Alch{\'{e}}{-}Buc}, \bibinfo{person}{Emily~B. Fox}, {and}
  \bibinfo{person}{Roman Garnett}} (Eds.). \bibinfo{pages}{6559--6569}.
\newblock
\urldef\tempurl%
\url{https://proceedings.neurips.cc/paper/2019/hash/e52ad5c9f751f599492b4f087ed7ecfc-Abstract.html}
\showURL{%
\tempurl}


\bibitem[\protect\citeauthoryear{Wilcoxon}{Wilcoxon}{1992}]%
        {wilcoxon1992individual}
\bibfield{author}{\bibinfo{person}{Frank Wilcoxon}.}
  \bibinfo{year}{1992}\natexlab{}.
\newblock \showarticletitle{Individual comparisons by ranking methods}.
\newblock In \bibinfo{booktitle}{\emph{Breakthroughs in statistics}}.
  \bibinfo{publisher}{Springer}, \bibinfo{pages}{196--202}.
\newblock


\bibitem[\protect\citeauthoryear{Wu, Schuster, Chen, Le, Norouzi, Macherey,
  Krikun, Cao, Gao, Macherey, Klingner, Shah, Johnson, Liu, Kaiser, Gouws,
  Kato, Kudo, Kazawa, Stevens, Kurian, Patil, Wang, Young, Smith, Riesa,
  Rudnick, Vinyals, Corrado, Hughes, and Dean}{Wu et~al\mbox{.}}{2016}]%
        {wu2016google}
\bibfield{author}{\bibinfo{person}{Yonghui Wu}, \bibinfo{person}{Mike
  Schuster}, \bibinfo{person}{Zhifeng Chen}, \bibinfo{person}{Quoc~V. Le},
  \bibinfo{person}{Mohammad Norouzi}, \bibinfo{person}{Wolfgang Macherey},
  \bibinfo{person}{Maxim Krikun}, \bibinfo{person}{Yuan Cao},
  \bibinfo{person}{Qin Gao}, \bibinfo{person}{Klaus Macherey},
  \bibinfo{person}{Jeff Klingner}, \bibinfo{person}{Apurva Shah},
  \bibinfo{person}{Melvin Johnson}, \bibinfo{person}{Xiaobing Liu},
  \bibinfo{person}{Lukasz Kaiser}, \bibinfo{person}{Stephan Gouws},
  \bibinfo{person}{Yoshikiyo Kato}, \bibinfo{person}{Taku Kudo},
  \bibinfo{person}{Hideto Kazawa}, \bibinfo{person}{Keith Stevens},
  \bibinfo{person}{George Kurian}, \bibinfo{person}{Nishant Patil},
  \bibinfo{person}{Wei Wang}, \bibinfo{person}{Cliff Young},
  \bibinfo{person}{Jason Smith}, \bibinfo{person}{Jason Riesa},
  \bibinfo{person}{Alex Rudnick}, \bibinfo{person}{Oriol Vinyals},
  \bibinfo{person}{Greg Corrado}, \bibinfo{person}{Macduff Hughes}, {and}
  \bibinfo{person}{Jeffrey Dean}.} \bibinfo{year}{2016}\natexlab{}.
\newblock \showarticletitle{Google's Neural Machine Translation System:
  Bridging the Gap between Human and Machine Translation}.
\newblock \bibinfo{journal}{\emph{CoRR}}  \bibinfo{volume}{abs/1609.08144}
  (\bibinfo{year}{2016}).
\newblock
\showeprint[arxiv]{1609.08144}
\urldef\tempurl%
\url{http://arxiv.org/abs/1609.08144}
\showURL{%
\tempurl}


\bibitem[\protect\citeauthoryear{Zhang, Wang, Zhang, Sun, Wang, and Liu}{Zhang
  et~al\mbox{.}}{2019}]%
        {zhang2019novel}
\bibfield{author}{\bibinfo{person}{Jian Zhang}, \bibinfo{person}{Xu Wang},
  \bibinfo{person}{Hongyu Zhang}, \bibinfo{person}{Hailong Sun},
  \bibinfo{person}{Kaixuan Wang}, {and} \bibinfo{person}{Xudong Liu}.}
  \bibinfo{year}{2019}\natexlab{}.
\newblock \showarticletitle{A novel neural source code representation based on
  abstract syntax tree}. In \bibinfo{booktitle}{\emph{Proceedings of the 41st
  International Conference on Software Engineering, {ICSE} 2019, Montreal, QC,
  Canada, May 25-31, 2019}}, \bibfield{editor}{\bibinfo{person}{Joanne~M.
  Atlee}, \bibinfo{person}{Tevfik Bultan}, {and} \bibinfo{person}{Jon Whittle}}
  (Eds.). \bibinfo{publisher}{{IEEE} / {ACM}}, \bibinfo{pages}{783--794}.
\newblock
\urldef\tempurl%
\url{https://doi.org/10.1109/ICSE.2019.00086}
\showDOI{\tempurl}


\bibitem[\protect\citeauthoryear{Zhang, Xie, Ma, Du, Hu, Liu, Zhao, and
  Sun}{Zhang et~al\mbox{.}}{2020}]%
        {zhang2020towards}
\bibfield{author}{\bibinfo{person}{Xiyue Zhang}, \bibinfo{person}{Xiaofei Xie},
  \bibinfo{person}{Lei Ma}, \bibinfo{person}{Xiaoning Du},
  \bibinfo{person}{Qiang Hu}, \bibinfo{person}{Yang Liu},
  \bibinfo{person}{Jianjun Zhao}, {and} \bibinfo{person}{Meng Sun}.}
  \bibinfo{year}{2020}\natexlab{}.
\newblock \showarticletitle{Towards characterizing adversarial defects of deep
  learning software from the lens of uncertainty}. In
  \bibinfo{booktitle}{\emph{2020 IEEE/ACM 42nd International Conference on
  Software Engineering (ICSE)}}. IEEE, \bibinfo{pages}{739--751}.
\newblock


\bibitem[\protect\citeauthoryear{Z{\"u}gner, Kirschstein, Catasta, Leskovec,
  and G{\"u}nnemann}{Z{\"u}gner et~al\mbox{.}}{2021}]%
        {Daniel21Language}
\bibfield{author}{\bibinfo{person}{Daniel Z{\"u}gner}, \bibinfo{person}{Tobias
  Kirschstein}, \bibinfo{person}{Michele Catasta}, \bibinfo{person}{Jure
  Leskovec}, {and} \bibinfo{person}{Stephan G{\"u}nnemann}.}
  \bibinfo{year}{2021}\natexlab{}.
\newblock \showarticletitle{Language-Agnostic Representation Learning of Source
  Code from Structure and Context}. In \bibinfo{booktitle}{\emph{{ICLR}
  (Poster)}}. \bibinfo{publisher}{OpenReview.net}.
\newblock


\end{thebibliography}

% \appendix






\end{document}